%% file: mnras_main.tex
\documentclass[fleqn,usenatbib]{mnras}

\usepackage{newtxtext,newtxmath}

\usepackage[T1]{fontenc}

\DeclareRobustCommand{\VAN}[3]{#2}
\let\VANthebibliography\thebibliography
\def\thebibliography{\DeclareRobustCommand{\VAN}[3]{##3}\VANthebibliography}

\usepackage{graphicx}	
\usepackage{amsmath}	
\usepackage{xcolor}
\usepackage{import}

\newcommand{\lya}{Ly$\alpha$}

\newcommand{\halpha}{\mbox{H$\alpha$}}
\newcommand{\hbeta}{\mbox{H$\beta$}}

\newcommand{\hI}{\mbox{H{\sc i}}}

\newcommand{\oII}{\mbox{O{\sc ii}}}
\newcommand{\oIII}{\mbox{O{\sc iii}}}

\newcommand{\msun}{\mbox{M$_\odot$}}

\newcommand{\galone}{J0004}
\newcommand{\galtwo}{J0139}
\newcommand{\galthree}{J0156}
\newcommand{\galfour}{J0232}
\newcommand{\galfive}{J2318}
\newcommand{\galsix}{J2353}
\newcommand{\galseven}{J2359}

\title[Low-$z$ \lya\ halos]{On the evolution of the size of Lyman alpha halos across cosmic time: no change in the circumgalactic gas distribution when probed by line emission}

\author[A. Runnholm et al.]{
Axel Runnholm$^{1}$,\thanks{axel.runnholm@astro.su.se}
Matthew J. Hayes$^{1}$,
Yu-Heng Lin$^{2}$,
Jens Melinder$^{1}$,
Claudia Scarlata$^{2}$,
Angela Adamo$^{1}$,
\newauthor
Ramona Augustin$^{3}$
Arjan Bik$^{1}$,
Jérémy Blaizot$^{4}$,
John M. Cannon$^{5}$,
Sebastiano Cantalupo$^{6,7}$,
Thibault Garel$^{8}$,
\newauthor
Max Gronke$^{9}$,
Edmund C. Herenz$^{10}$,
Floriane Leclercq$^{11}$,
Göran Östlin$^{1}$,
Celine Peroux$^{12,13}$,
\newauthor
Armin Rasekh$^{1}$,
Michael J. Rutkowski$^{14}$,
Anne Verhamme$^{8}$, and
Lutz Wisotzki$^{15}$
\\
$^{1}$Department of Astronomy, Oscar Klein Centre, Stockholm University, AlbaNova universitetscentrum, SE-106 91 Stockholm, Sweden\\
$^{2}$Minnesota Institute for Astrophysics, School of Physics and Astronomy, University of Minnesota, 316 Church str SE, Minneapolis, MN 55455,USA\\
$^{3}$Space Telescope Science Institute, 3700 San Martin Drive, Baltimore, MD, 21218, USA\\
$^4$Univ Lyon, Univ Lyon1, Ens de Lyon, CNRS, Centre de Recherche Astrophysique de Lyon UMR5574, F-69230, Saint-Genis-Laval, France\\
$^5$Department of Physics and Astronomy, Macalester College, 1600 Grand Avenue, Saint Paul, MN 55105, USA\\
$^6$Dipartimento di Fisica, Universitá di Milano Bicocca, Piazza della Scienza 3, 20126 Milano, Italy\\
$^7$Institute for Astronomy, ETH Zurich, 8093 Zurich, Switzerland\\
$^8$Department of Astronomy, University of Geneva, Chemin Pegasi 51, 1290 Versoix, Switzerland\\
$^9$Max Planck Institut fur Astrophysik, Karl-Schwarzschild-Straße 1, D-85748 Garching bei München, Germany\\
$^{10}$European Southern Observatory, Av. Alonso de C\'ordova 3107, 763 0355 Vitacura,
Santiago, Chile\\
$^{11}$Department of Astronomy, The University of Texas at Austin, 2515 Speedway, Stop C1400, Austin, TX 78712, USA\\
$^{12}$European Southern Observatory, Karl-Schwarzschildstrasse 2, D-85748 Garching bei M{\"u}nchen, Germany\\
$^{13}$Aix Marseille Universit\'e, CNRS, LAM (Laboratoire d'Astrophysique de Marseille) UMR 7326, 13388, Marseille, France\\
$^{14}$Department of Physics and Astronomy, Minnesota State University, Mankato, MN 56001, USA\\
$^{15}$Leibniz-Institut für Astrophysik Potsdam (AIP), An der Sternwarte 16, 14482 Potsdam, Germany\\
}

\date{Accepted XXX. Received YYY; in original form ZZZ}

\pubyear{2015}

\begin{document}
\label{firstpage}
\pagerange{\pageref{firstpage}--\pageref{lastpage}}
\maketitle

\begin{abstract}
Lyman $\alpha$ (\lya) is now routinely used as a tool for studying  high-redshift galaxies and its resonant nature means it can trace neutral hydrogen around star-forming galaxies. Integral field spectrograph measurements of high-redshift \lya\ emitters indicate that significant extended \lya\ halo emission is ubiquitous around such objects. We present a sample of redshift 0.23 to 0.31 galaxies observed with the Hubble Space Telescope selected to match the star formation properties of high-$z$  samples while optimizing the observations for detection of low surface brightness \lya\ emission. 
The \lya\ escape fractions range between 0.7\% and 37\%, and we detect extended \lya\ emission around six out of seven targets.
We find \lya\ halo to UV scale length ratios around 6:1 which is marginally lower than high-redshift observations, and halo flux fractions between 60\% and 85\% ---consistent with high-redshift observations---when using comparable methods. However, our targets show additional extended stellar UV emission: we parametrize this with a new double exponential model. We find that this parametrization does not strongly affect the observed \lya\ halo fractions. 
We find that deeper \halpha\ data would be required to firmly determine the origin of \lya\ halo emission, however, there are indications that \halpha\ is more extended than the central FUV profile, potentially indicating conditions favorable for the escape of ionizing radiation. 
We discuss our results in the context of high-redshift galaxies, cosmological simulations, evolutionary studies of the circumgalactic medium in emission, and the emission of ionizing radiation. 
\end{abstract}



\input{introduction}

\input{data}

\input{methods}

\input{results}

\input{discussion}

\input{conclusions}

\section*{Acknowledgments}
This work is based on observations made with the NASA/ESA Hubble Space Telescope, obtained at the Space Telescope Science Institute, which is operated by the Association of Universities for Research in Astronomy, Inc., under NASA contract NAS 5-26555. These observations are associated with program 15643
M.H. is supported by the Knut \& Alice Wallenberg Foundation. 
RA was supported by NASA grant 80NSSC18K110.

\section*{Data Availability}
The data underlying this article will be shared on reasonable request to the corresponding author.
%




\bibliography{Halos}{}
\bibliographystyle{mnras}


\bsp	
\label{lastpage}
\end{document}

%% file: introduction.tex
\section{Introduction}\label{introduction}

Lyman \(\alpha\) (\lya) results from the transition from the \(2p\) to \(1s\) energy levels of hydrogen and, under normal gas conditions, 68\% of ionizing photons are processed through this transition \citep{dijkstra2019}. This, coupled with the prevalence of hydrogen gas in galaxies, means that it is intrinsically the strongest emission line of star-forming systems. However, since this transition is from the ground state of hydrogen, most astrophysical media containing even small amounts of neutral hydrogen are optically thick to \lya.

The resonant nature of the \lya\ transition means that \lya-photons absorbed by hydrogen are re-emitted in that same transition, and scatter on neutral hydrogen. This has the interesting and very important consequence that some properties of the gas in an emitting galaxy can be encoded in the \lya\ radiation.

If we can understand how the spatial profiles of \lya\ relate to the properties of the galaxy as a whole, it could improve our understanding of how the neutral gas is distributed in and around such systems. Since \lya\ is in many cases very bright, we can potentially spatially map neutral hydrogen at much larger redshifts than what is practically observable using direct tracers such as 21 cm emission \citep[see e.g.][]{obreschkow2011}. This is especially interesting for studying the circumgalactic (CGM) and intergalactic media (IGM). This tenuous gas can typically only be studied on fortuitous sightlines where the CGM is intersected by a bright background continuum source. \lya\ imaging, on the other hand, could be used to map the distribution of gas around an individual galaxy.

This idea was first explored using narrowband imaging of Lyman Break Galaxies (LBGs) by \citet{steidel2011} where stacking of sources allowed detection of low surface brightness emission. These results indicated significant and very extended  \lya\ emission around the median stack with a characteristic scale length of 25 kpc. More recent results using integral field spectrographs (IFSs), such as the Multi-Unit Spectroscopic Explorer \citep[MUSE, ][]{Bacon.2010} on the Very Large Telescope (VLT) and the Keck Cosmic Web Imager \citep[KCWI, ][]{morrissey2018} on Keck, have clearly detected \lya\ halo emission around high-\(z\) galaxies. Using sensitive IFSs allows the detection of individual halos \citep[see e.g.][]{erb2018a}. These halos are indeed very extended, with exponential scale lengths between 1 and 18 kpc with a median around 4.5 kpc \citep{wisotzki2016,leclercq2017, leclercq2020}, which is much larger than the detected UV emission of these galaxies. \citet{kusakabe2022} studied \lya\ emission around UV selected galaxies, also using MUSE, and found an incidence rate of 80\% and that the halos extend as far as 40 kpc. Scale lengths similar to the MUSE results have also  been found in narrowband studies including one subsample of \citet{feldmeier2013} and the results of \citet{momose2014, momose2016}. \citet{chen2021} used KCWI and found more extended \lya--halos with a scale length of the median stack of 17.5 kpc, more similar to the LBGs studied by \citet{steidel2011}. \citet{niemeyer2022} present the median-stacked \lya\ surface brightness profiles of 968 spectroscopically selected \lya\ emitting galaxies (LAEs) at redshifts $1.9 < z < 3.5$ in the early data of the Hobby-Eberly Telescope Dark Energy Experiment (HETDEX). They find that the median-stacked radial profile at r $< 80$ kpc agrees with the results from the MUSE sample, but is more extended at r $> 80$ kpc. These instruments have opened up a new frontier of research where we can not only study individual \lya\ halos, but samples of individual halo detections can become large enough for population statistics to be derived. 

Despite the ease with which such high-$z$ surveys can be conducted, they remain fundamentally limited by the lack of ancillary data, such as rest-frame optical spectroscopy and high-resolution imaging of the stellar population, that can be used to infer the properties of the systems that give rise to the \lya\ halos. In order to study the origin of \lya\ halos and the properties of their host galaxies, we need to turn to observations in the low redshift universe.

However, studying \lya\ at low redshift is challenging -- primarily because the UV wavelength of \lya\ requires space based observations. Early low redshift studies such as \citet{kunth2003a}, \citet{hayes2005a,hayes2007} and \citet{ostlin2009} used a \lya\ transmitting narrowband filter on the Solar Blind Channel (SBC on the Hubble Space Telescope (HST) Advanced Camera for Surveys (ACS). However, \citet{hayes2009b} showed that synthesising narrowband observations from adjacent long pass broad band observations was more effective. The Lyman Alpha Reference Sample (LARS) \citep{ostlin2014,hayes2014} has successfully used this method to image \lya, and has shown that there are significant differences in the morphologies of \lya\ compared to that of \halpha\ and UV. The high spatial resolution of LARS enables the study of small structural differences and has shown, for instance, that knots and structures present in UV and \halpha\ are absent in \lya\ and that the \lya\ structure, in general, appears smoother \citep{bridge2017a}. The centroids of \lya\ emission and its position angle also significantly differ from the stellar UV emission \citep{rasekh2021}. 

LARS, which consists of galaxies at redshift \(\lesssim0.2\), has confirmed significant \lya~halos extending 1 to 4 times further than both \halpha~and stellar far UV (FUV) \citep{hayes2013, rasekh2021}. \citet{yang2017b} used a different technique based on COS acquisition images and found very similar ratios with \lya\ extending 2-4 times further than the UV. There is some indication that high-redshift galaxies such as those observed by \citet{leclercq2017} have larger ratios than this, hinting that there may be some redshift evolution in halo sizes. However, neither \citet{wisotzki2016} nor  \citet{leclercq2017} see any redshift evolution within their samples despite the significant redshift range (between $z\sim3$ and $z\sim6.6$) it covers. It is important to note, however, that this redshift range 
corresponds to only 1.3 Gyr, in contrast to the $\sim$11 Gyr that elapses between $z=3$ and 0.

Determining whether there is a change in halo sizes between low and high $z$ is crucial for understanding galaxy evolution, since it would imply a significant physical evolution in the properties of \lya-halo-hosting galaxies over this time. Absorption line studies, such as COS-Halos \citep{tumlinson2013}, COS-GASS \citep{borthakur2015} and COS-Bursts \citep{heckman2017}, have shown that the atomic gas halos of star forming galaxies extend far beyond the stellar components with absorption equivalent widths (EWs) higher than 0.1 Å (corresponding to gas column densities $\gtrsim10^{17}$ which, for normal gas temperatures, implies $\tau \gtrsim 1000$ for \lya) as far as 300 kpc from the central galaxy. This appears to hold regardless of viewing angles of the galaxies, indicating a covering fraction of 1 for this neutral medium. The CGM of $z\approx0$ galaxies therefore do have sufficient \hI\ in their halos to scatter \lya\ out to large distances and yet the observed \lya\ emission around individual galaxies declines significantly faster than the neutral gas column does.

LARS is optimized to study the detailed structural differences between FUV, \halpha, and \lya, and therefore selects very low redshift targets to get the best possible spatial resolution. The field of view (FoV) corresponds to $\sim10$ kpc at the redshift of LARS \citep{rasekh2021}. This means that the observations cannot cover radii that come even close to those probed by the absorption line studies mentioned above and can only probe $\sim2$ scale lengths of a median \lya\ halo ($\sim4.5$ kpc) found in \citet{leclercq2017}. Consequently, the observations are more optimized to characterize the detailed morphology of the \lya\ emission and absorption rather than the low surface brightness extent of them. In order to probe larger scales, and definitively determine whether \lya\ halos evolve between $z\sim0$ and $z>3$, a larger FoV is required.  Therefore, we designed an HST program for doing LARS type imaging at $0.23 \leq z \leq 0.31$ where we can obtain such observations with minimal background from geocoronal emission. In this work we present the results of this HST program.

The paper is structured as follows: in Section\,\ref{sec:Data} we present the data set and the details of our HST observations. In Section\,\ref{methods} we discuss the methodology, with particular focus spent on the data reduction and production of \lya~images. In Section\,\ref{results} we present and discuss the primary results from the \lya\ imaging and in Section\,\ref{sec: sample discussion} we give detailed descriptions of the objects in this study. Finally we discuss the potential implications of our findings  in Section\,\ref{Discussion} and conclude in Section\,\ref{Conclusions}. We assume a flat $\Lambda$CDM cosmology with H$_0 = 70$ kms$^{-1}$Mpc$^{-1}$, $\Omega_m = 0.3$, $\Omega_\Lambda = 0.7$.

%% file: data.tex
\section{Data}\label{sec:Data}

\begin{figure*}
    \centering
    \includegraphics[width=\textwidth]{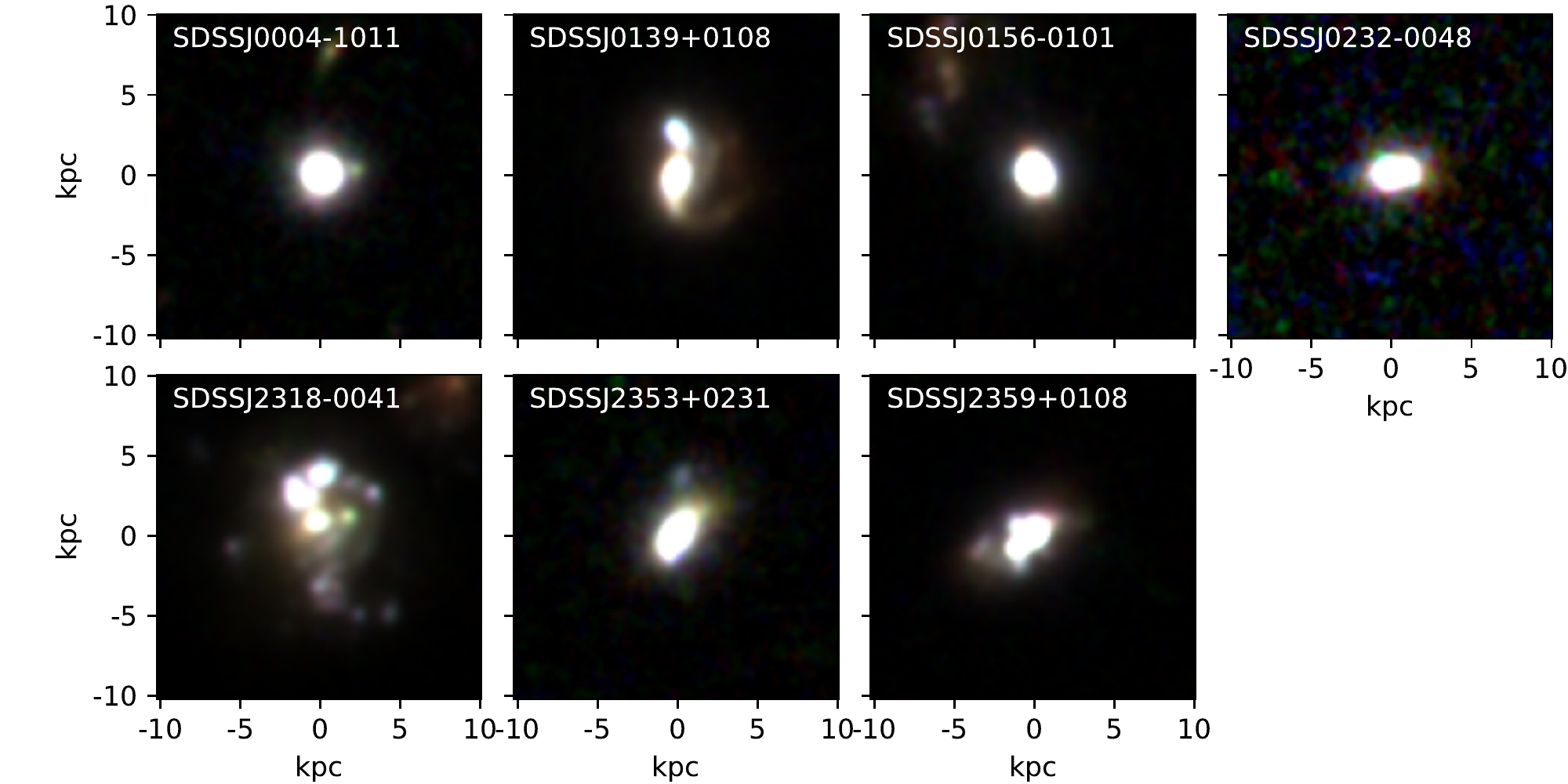}
    \caption{Optical 3-color composite images of the seven targets. Filter setup: Red - F814W, Green - F555W, Blue - F435W}
    \label{fig:optical images}
\end{figure*}

We obtained HST imaging observations in six bands of seven galaxies at redshifts $0.23 \leq z \leq 0.31$ (see Table\,\ref{tab: galaxy properties} for details) taken as part of GO 15643, PI Hayes. Throughout the paper we will refer to the targets using the short name introduced in this table. HST/COS spectral observations of these targets will be presented in a future paper. The HST filters included two UV filters in the Solar Blind Channel (SBC) of ACS (F150LP, and F165LP), that were selected to enable synthetic \lya\ narrowband observations (see section \ref{sec:modelling-the-continuum} for details on the continuum subtraction of the \lya\ images).  In addition we used three optical filters in the Wide Field Channel of ACS (WFC; F435W, F555W, F814W) and a narrowband (FR782N or FR853N depending on redshift) to capture \halpha, also in ACS/WFC. The observations are summarized in Table\,\ref{tab: summary of observations} and optical composites showing the morphologies of the targets are displayed in Figure\,\ref{fig:optical images}.

\begin{table}
\caption{\label{tab: galaxy properties} The sample of galaxies} 

\begin{tabular}{lllll}

Galaxy& 
Short name&
RA$^*$& 
Dec$^*$& 
Redshift\\
\hline
SDSSJ0004$-$1011 & \galone\   &  00 04 30.3 & $-$10 11 29.6 &  $0.2386$\\ 
SDSSJ0139$+$0108 & \galtwo\   &  01 39 13.2 & $+$01 08 56.0 &  $0.3073$\\ 
SDSSJ0156$-$0101 & \galthree\ &  01 56 55.8 & $-$01 01 16.5 & $0.2696$\\ 
SDSSJ0232$-$0048 & \galfour\  &  02 32 43.6 & $-$00 48 32.3 & $0.3095$\\ 
SDSSJ2318$-$0041 & \galfive\  &  23 18 13.0 & $-$00 41 26.0 & $0.2517$\\ 
SDSSJ2353$+$0231 & \galsix\   &  23 53 35.5 &  $+$02 31 50.2 &  $0.2333$\\ 
SDSSJ2359$+$0108 & \galseven\ &  23 59 26.7 &  $+$01 08 38.8 &  $0.2607$\\ 
\hline 
\end{tabular}
\hbox{$^*$ RA and Dec are in J2000}
\end{table}

\begin{table*}
\caption{HST exposure times in seconds\label{tab: summary of observations}}
\begin{tabular}{ccccccc}
\hline
\emph{Bandpass alias} & \lya & FUV continuum & U$^*$ & B$^*$ & I$^*$ & H$\alpha$ \\
\emph{HST Instrument} & ACS/SBC & ACS/SBC & ACS/WFC & ACS/WFC & ACS/WFC & ACS/WFC \\
\emph{HST Filter name} & F150LP & F165LP & F435W & F555W & F814W &  \\
Galaxy       &        &        &       &       &       &  \\
\hline
\galone\   & 5104 & 7698 & 1132 & 1030 & 800 &  1300 (FR782N)\\
\galtwo\   & 5168 & 7600 & 1116 & 1016 & 778 &  1300 (FR853N)\\
\galthree\ & 5088 & 7680 & 1116 & 1016 & 778 &  1300 (FR853N)\\
\galfour\  & 5108 & 7660 & 1116 & 1016 & 778 &  1300 (FR853N)\\
\galfive\  &5150 & 7751 & 1116 & 1016 & 778 &  1300 (FR853N)\\
\galsix\   & 5090 & 7678 & 1128 & 1026 & 794 &  1300 (FR782N)\\
\galseven\ & 5078 & 7690 & 1116 & 1016 & 778 &  1300 (FR853N)\\
\hline
\end{tabular}

\hbox{$^*$ These aliases refer to the approximate restframe bandpasses.}

\end{table*}

\subsection{Sample selection }\label{sec:sample selection}

\begin{figure*}
\includegraphics[width=\textwidth]{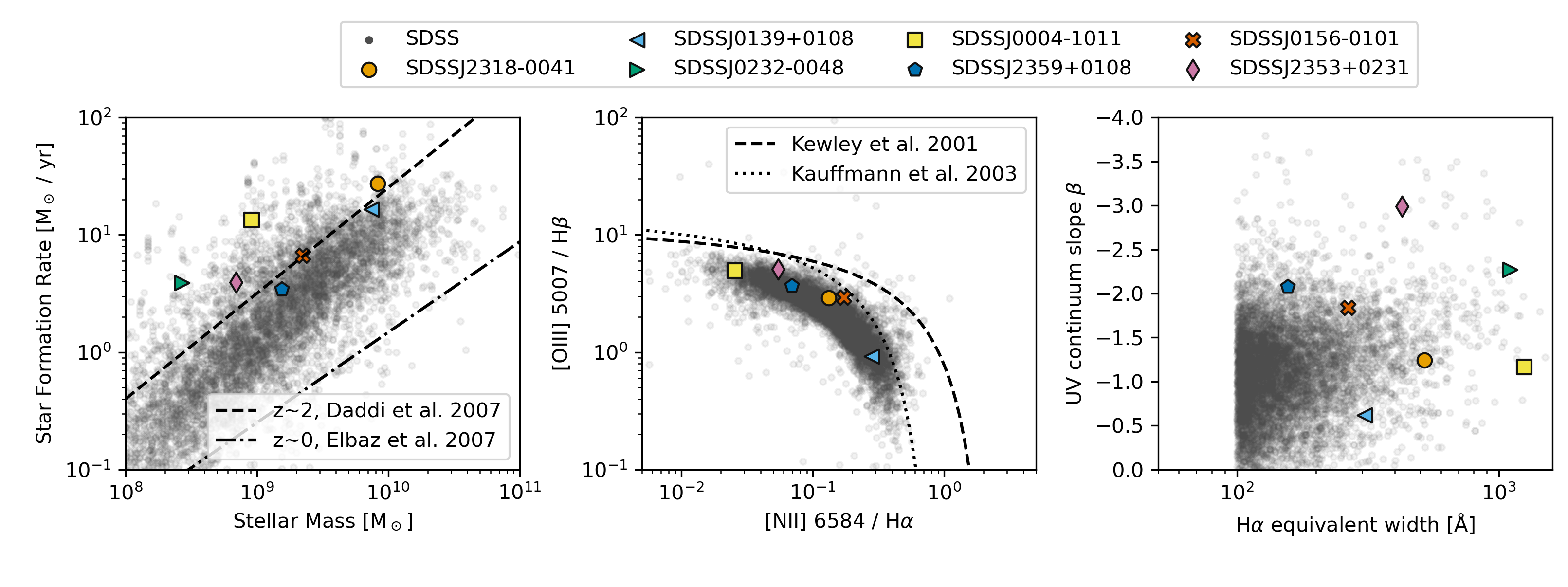}
\caption{Sample selection figures. Left panel shows star formation rate against stellar mass. The black points show a sample of 7000 SDSS galaxies with the single selection criterion of having high ($\geq100$ \AA) \halpha\ equivalent width. The dash-dotted line shows the star forming main sequence relation at redshift $\sim0$  derived from the SDSS from \citet{elbaz2007} and the dashed line shows the main sequence at redshift 2 from \citet{daddi2007}. The central panel shows the BPT diagram \citep{baldwin1981} with the relations from \citet{kewley2001} and \citet{kauffmann2003} showing the demarcation between star forming galaxies and AGN. The third panel shows the UV slope $\beta$  (assuming the standard proportionality $f_\lambda\propto \lambda ^{\beta}$) derived from \emph{GALEX} photometry against the \halpha\ equivalent width.}
\label{fig:selection}
\end{figure*}

We select objects with redshift such that \lya\ falls in the F150LP but not the F165LP filter. This limits the redshift range to $0.23<z< 0.31$. Compared to most previous studies, this filter combination ensures that no strong skylines, specifically OI at 1305 and 1356~\AA, are transmitted by the filters. Over this redshift range, the ACS/SBC detector covers a physical distance in excess of $\approx60$ kpc.  This ensures that there is sufficient space in the frame to subtract any remaining background from the image while being confident that no \lya\ signal is lost in the process.

Sources were selected from a catalog of re-fitted SDSS DR 14 \citep{abolfathi2018} spectra according to the following criteria:

\begin{enumerate}
\item
  Less than 0.2 magnitudes $u$-band  extinction from the Milky Way
\item
  Declination below +5 degrees to ensure visibility from Cerro Paranal to enable further follow-up studies
\item
  Full Width at Half Maximum (FWHM) of the optical emission lines less than 250 km s$^{-1}$ and a position in the BPT diagram below the \citet{kauffmann2003} demarcation line in order to ensure that contamination from AGN was kept to a minimum
\item
  Far UV absolute magnitude $M_\mathrm{FUV}$ in the range $-17.5$ to $-22$ from GALEX \citep{martin2003} in order to overlap with high-redshift galaxies. Specifically the lower end was selected to match the 25th percentile of \citet{leclercq2017} and the upper end was selected to match the upper brightness limit of \citet{steidel2011}. 
\item
  \halpha\ equivalent width larger than 100 \AA\ in order to ensure that all galaxies produce significant \lya. This also enables direct comparison between this sample of galaxies and the LARS sample at lower redshift. 
\end{enumerate}

After these criteria were applied there were still more than 200 galaxies that could be included in the sample. The seven galaxies were then selected to span the full range of galaxy properties in terms of position on the star forming main sequence, BPT diagram, and UV slope versus \halpha\ equivalent width (see Figure\,\ref{fig:selection}).

In the first panel of Figure~\ref{fig:selection} the dashed line shows the star forming main sequence at redshift $\sim2$ from \citet{daddi2007} and our seven galaxies lie close to and in general above that relation, demonstrating that they are, at least in this respect, comparable to high-redshift galaxies.

%% file: methods.tex
\section{Methods}\label{methods}
\subsection{Observations and Data reduction}\label{sec:data-reduction}

	The total exposure times and filters used are summarized in Table\,\ref{tab: summary of observations}. The observations were performed with  a custom calculated large ($\sim3$ arcsec) dithering pattern specially designed to minimize flat field uncertainties and biases, as well as allow for sub-pixel sampling of the PSF.

\subsubsection{UV data}\label{uv-data}

	The individual UV exposures (from the ACS/SBC) were initially reduced using the CALACS pipeline with the latest calibration files to produce flatfield-corrected ({\sc flt}) frames.  We implemented an additional, custom step to correct for residual dark-current in the SBC detector. This dark-current is strongly temperature dependent and therefore varies between exposures and time on orbit. The procedure used a collection of dark-current images taken at various times and temperatures, and fits the dark-current image to each SBC frame with $\chi^2$ minimization of the form
	\begin{equation}
	B = A \times I_{\mathrm{DC}} + C    
	\end{equation}
	where $B$ is the total fitted background, $A$ is the amplitude which is allowed to vary, $I_{\mathrm{DC}}$ is a given dark-current image, and C is a constant that represents the actual contribution of (flat) sky background. The best fit background and dark-current was then subtracted from each {\sc flt} frame.

	This procedure works very well for all of the data that were taken as a part of GO 15643. However, two of our F150LP exposures are archival data taken as a part of GO 11107 (PI Heckman) and these images showed an additional background gradient after the dark-current subtraction. In these cases we mask the edges of the image as well as the source itself by selecting a broad circular annulus centered on the galaxy and fit an additional plane to the flux in this region which is then subtracted.

\subsubsection{Optical data}\label{se:optical-data}

	The optical data (from ACS/WFC) were pipeline-processed in the same manner as the UV data. Since each optical filter only contains two exposures we found that drizzling could not adequately remove all cosmic rays from the frames. We therefore did a separate cosmic ray rejection on the frames using the \texttt{astroscrappy} tool \citep{mccully2018} which is based on \texttt{lacosmic} \citep{vandokkum2001} before drizzling.

	The H$\alpha$ narrowband images were taken using tunable ramp filters. This allows us to center the band on the \halpha\ line, but those filters do not cover the whole chip. This means that standard \texttt{astrodrizzle} background estimates are inaccurate. We therefore instead fit and subtract the background of the unvignetted region using a flat plane in the same way as was done for the GO 11107 UV images.

\subsubsection{Image alignment}\label{image-alignment}

	After the pipeline reduction and dark-current subtraction, we performed fine-alignment of the images. Due to the lack of stars in the fields, especially in the FUV filters, this could not be done using the standard \texttt{TweakReg} task from \texttt{astrodrizzle}. We therefore developed a custom cross-correlation based alignment routine.  

	The first step was to make sure that all exposures taken with a given filter were well aligned with each other. For the optical filters, only one HST visit with two individual exposures was used, which meant that the default pipeline World Coordinate Systems (WCS) were in general well matched. In two visits this was not the case however: F555W for \galtwo\ and FR853N for \galseven. These were treated according to the methodology developed for the UV filters (see below).

	For the UV filters there were 3 individual visits per galaxy per filter, and in most cases even the fine alignment of exposures belonging to the same visit was poor (offsets $\gg1$px). We therefore had to align all exposures in each filter to each other. This was done as follows:

	\begin{enumerate}
	\item
	  Each exposure was individually drizzled to ensure that all images had the same pixel scale and were derotated. We used a pixel scale of 0\farcs01 at this stage to ensure that alignment could be done on a sub-pixel level compared to the final pixel scale (0\farcs04).
	\item
	  The exposures were then compared to each other using a 2 dimensional
	  correlation analysis (using \texttt{scipy.signal.2dcorrelation}), and the point of maximum correlation was converted into a pixel shift.
	\item
	  These pixel shifts were then converted into right ascension and
	  declination, and added as a shift to the  WCS of the original images.
	\end{enumerate}

	The philosophy of this approach is to mimic that of \texttt{tweakreg} in \texttt{astrodrizzle}, which minimizes the number of times pixel data are resampled, compared to manual realignment and regridding post-drizzle, and to maximize the correlation of signal between pixels in the final images. 
	For \galtwo\ and \galfour\ the individual F165LP exposures had insufficient signal to noise to produce stable correlation analysis. In these cases we instead drizzled the exposures in each visit together. These visit-level drizzles were then aligned to each other as described above.

	We then drizzle each filter together, producing in total 1 frame per filter, again using a small pixel size of 0.01\arcsec. These frames are then used as input to an inter filter correlation analysis where the F435W frame was used as the reference. Again, the resulting shifts are converted to \(\Delta\)RA and \(\Delta\)Dec and added to the WCS of the original frames. This alignment process results in six individual drizzled images that are aligned onto a common grid with consistent world-coordinate systems. The final pixel scale used was 0\farcs04. 

\subsubsection{Re-estimating the UV uncertainties}
	A simple comparison between the RMS values in the sky regions of the final drizzled image and compared to the median of the errors given in the weight frame produced by drizzle indicated that the drizzle errors were significant overestimates of the true standard deviations, by factors of up to 10. We therefore re-estimate the error as the quadrature sum of a constant RMS estimated from the sky regions of the frame and the square root of the signal (counts) in the frame. Additionally, the drizzling means that the noise in the image is correlated. We therefore use the correction factor given in \citet{fruchter2002} to correct for this:
	\begin{equation}
	R = \frac{1}{1 - \frac{1}{3r}}
	\end{equation}
	where r = pixfrac / scale and the pixfrac is set to 1 in this case. This expression is valid for a filled uniform dither. Our observations were designed for this to be a reasonable approximation. Within each visit there were several dithering positions taken, and each one was designed to reconstruct the PSF with 1/3 pixel dithers in x and y.

\subsubsection{PSF Matching}\label{sec: psf-matching}
	The Point Spread Functions (PSF) of the HST cameras varies strongly between the ACS/SBC and the ACS/WFC, and also between the different filters in each camera.  After the images are aligned, we then match images obtained in each filter to a common PSF because our modeling requires spectral analysis to be performed at the level of individual pixels.  

	For the optical filters we construct PSFs based on TinyTim modeling \citep{krist2011}. The raw PSF models from TinyTim are added to the individual science frames and the final models are then extracted from the inserted PSFs in the drizzled and stacked images. For WFC data where spatial variance of the PSF does not need to be taken into account, TinyTim provides an accurate characterization of the PSF, however the situation is more complex for the ACS SBC. The PSF of this instrument is characterized by a narrow core and strong wings which are not captured in the TinyTim models. See for instance \citet{hayes2016} and \citet{Melinder2023} for  an extensive discussion of this. For this reason we have created a fully empirical PSF model for the SBC filters based on observations of single stars in the globular cluster NGC 6681.  Based on the models of the individual PSFs, we make and apply a convolution kernel that homogenizes the images.

\subsubsection{Voronoi binning}\label{sec:voronoi binning}
	To ensure sufficient signal-to-noise for SED-fitting (which we use to model the continuum in the \lya\ transmitting filter, see Section\,\ref{sec:modelling-the-continuum}), the data were binned using a Voronoi tesselation algorithm. The algorithm used was the Weighted Voronoi Tesselation \citep{diehl2006} with the FUV image as a reference. The target signal-to-noise was 5 and the maximum bin size was set to 50 pixels (0.08 arcsec$^2$).

\subsubsection{Final background subtraction}
	Since we are interested in very low surface brightness emission, the quality of the background subtraction is crucial. Therefore, as a final step of the data reduction, we evaluated and removed any residual background in the drizzled images. We masked the data by selecting a wide annulus around the galaxies and fitted a plane which was then subtracted from the image, leaving a flat background centered on zero.

\subsection{Data analysis}
\subsubsection{Continuum subtracting \halpha}\label{sec:continuum subtraction halpha}
	In order to get a line-map of \halpha\ emission we need to subtract the stellar continuum contribution from the narrowband observations which we do by using the following expression \citep{hayes2006}:
	\begin{equation}
	F_{\mathrm{line}} = \frac{W_\mathrm{b}F_\mathrm{n} - W_\mathrm{n}F_\mathrm{b}}{W_\mathrm{b}W_\mathrm{n}}
	\label{eq:Fline}
	\end{equation}
	where \(F_\mathrm{b}\) and \(W_\mathrm{b}\) are the flux and width of
	the broadband -- F814W, and \(F_\mathrm{n}\) and \(W_\mathrm{n}\) are
	the flux and width of the narrowband. \(W_\mathrm{b}\) is taken to be
	1260.3 \AA~from the instrument manual. For the narrowband we calculate
	the width based on the expression given in \citet{bohlin2000}. We also estimate the uncertainties in this \halpha\ frame using a 500-iteration Monte Carlo simulation where we randomize the broadband and narrowband fluxes based on their error frames.

	\begin{table*}
	\caption{\label{tab: linefluxes} Measured optical line fluxes.} 
	\begin{tabular}{lllllllll}
	\hline
		Galaxy& 
		H$\alpha$& 
		H$\beta$& 
		H$\gamma$& 
		[OIII]$_{5007}$& 
		[OIII]$_{4363}$& 
		[OII]$_{3726,3729}$& 
		[NII]$_{6584}$& 
		[SII]$_{6717}$ \\
	\hline
	\galone\   &	$1420\pm90$ &	$490\pm31$ &	$245\pm19$ &	$2440\pm188$ &	$70\pm7$ & $740\pm50$ &	$36\pm6$ &	$57\pm6$\\ 
	\galtwo\   &	$990\pm130$ &	$345\pm45$ &	$170\pm30$ &	$320\pm50$ &	$14\pm13$ & $1100\pm175$ &	$280\pm30$ &	$165\pm33$\\ 
	\galthree\ &	$535\pm54$ &	$187\pm19$ &	$88\pm12$ &	$550\pm60$ &	$7.5\pm2$ & $520\pm60$ &	$92\pm9$ &	$60\pm6$\\ 
	\galfour\  &	$230\pm40$ &	$80\pm13$ &	$38\pm8$ &	$490\pm91$ &	$11\pm3$ &	$102\pm18$ &	$0.8\pm3.6$ &	$6.4\pm2.4$\\ 
	\galfive\  &	$2600\pm100$ &	$900\pm34$ &	$410\pm23$ &	$2600\pm116$ &	$31\pm9$ & $3000\pm130$ &	$340\pm12$ &	$295\pm11$\\ 
	\galsix\   &	$440\pm93$ &	$154\pm32$ &	$63\pm17$ &	$785\pm175$ &	$6.6\pm2.8$ & $330\pm76$ &	$24\pm5$ &	$24\pm6.7$\\ 
	\galseven\ &	$300\pm35$ &	$105\pm12$ &	$47\pm9$ &	$388\pm53$ &	$7\pm4$ & $390\pm55$ &	$21\pm8$ &	$40\pm8$\\ 
	\hline
	\end{tabular}
	\hbox{All fluxes are given in units of $10^{-17}$ erg s$^{-1}$ cm$^{-2}$ and corrected for foreground and internal dust extinction.}
	\end{table*}

	\subsubsection{Modelling the UV continuum}\label{sec:modelling-the-continuum}
	To estimate the stellar continuum at the wavelength of  \lya\ we use the \texttt{Lyman\ Alpha\ eXtraction\ Software} (\texttt{LaXs}; \citealt{hayes2009b}), the latest version of which is presented in detail in \citet{Melinder2023}. For completeness, we provide a summary of the modelling steps below:

	\texttt{LaXs}  models the SED in each pixel, or in this case Voronoi cell, and uses the models to estimate the continuum at \lya. The input to \texttt{LaXs} is therefore a set of images from all of our HST bands that have been drizzled to the same pixel scale and PSF matched. It then models the SED for each pixel. The SED model fits 2 stellar populations, one young and one old, and a dust extinction. The age of the old population is kept fixed to 1 Gyr, whereas the young population age is allowed to vary between 1 and 100 Myr. The stellar population spectra are taken from the spectral synthesis code Starburst99 \citep{leitherer1999a} and are matched to the nebular metallicity of the galaxy, as measured by the O3N2 metallicity indicator \citep[][see Table~\ref{tab: derived props}]{marino2013}, as closely as possible without interpolating between models. For these fits we also need to select a dust attenuation law---on the basis of the low metallicities and dwarf nature of our galaxies we opt for the SMC dust attenuation law \citep{prevot1984} in all cases. The SED fit is used to calculate a pixel-by-pixel scaling factor that is then applied to the FUV image to create the continuum map at the wavelength sampled by the F150LP filter. This estimated continuum is then subtracted from the F150LP image to create the final \lya\ image.

	We use a 500-realization Monte Carlo (MC) simulation to estimate the statistical errors in each pixel of the resulting \lya\ images. In each iteration the input images are perturbed according to their error frames and the fits are rerun. The full set of results of the MC are saved as a datacube which we use to estimate errors in our measurements (See Section\,\ref{sec: global props} and Section\,\ref{sec: fitting methodology} for specifics).

	In our case we give \texttt{LaXs} four continuum bands to fit: F165LP, F435W, F555W and F814W. However, the F814W also contains a contribution from the \halpha\ line. Therefore we first use the following companion result to equation\,\ref{eq:Fline} to calculate the line-corrected continuum flux density
	\begin{equation}
	  f_\mathrm{cont} = \frac{F_\mathrm{b} - F_\mathrm{n}}{W_\mathrm{b} - W_\mathrm{n}}
	\end{equation}
	The line-corrected F814W image is then used in the SED fit in place of the standard F814W data. 

	The ramp filters in ACS/WFC also transmit the [N~{\sc ii}]~$\lambda\lambda 6548,6584$~\AA\ doublet.  We correct for this contamination in a global sense using the spectroscopically measured [N~{\sc ii}] to \halpha\ line ratio (N2 index, see Section\,\ref{sec:optical spectroscopy}). 

\subsubsection{Optical Spectroscopy}\label{sec:optical spectroscopy}

	To measure the optical spectroscopic properties of our galaxies we downloaded the SDSS optical spectra and reanalyzed them. The fitting code that we use simultaneously measures the fluxes of 18 optical lines that are constrained to have a  common redshift, i.e. centroid shift. The code also takes into account the varying spectral resolution of the SDSS spectrograph by linearly interpolating the spectroscopic resolving power R which varies from 1600 at $\sim3000$ Å to 3500 at $\sim9000$ Å. The full linewidth of the spectral lines is then taken as a convolution of the width from the spectrograph and an intrinsic linewidth which is fitted for and kept common for all lines.

	We correct our linefluxes for dust absorption using the \texttt{PyNeb} \citep{luridiana2015} reddening correction module. We correct for both Milky Way extinction and internal dust attenuation using the CCM89 law \citep{cardelli1989}.  Note that when describing the SED fit to the stellar continuum above (Section~\ref{sec:modelling-the-continuum}) we used the SMC extinction law---the reason for the difference is that our spectral fitting aims to best estimate the behaviour of dust in the ultraviolet, whereas in the optical the SMC and Milky Way laws are almost indistinguishable.  The measured line fluxes are shown in Table\,\ref{tab: linefluxes} and derived quantities such as metallicity, ionization parameter, and dust extinction are shown in Table\,\ref{tab: derived props}.

	\begin{table} 
	\caption{\label{tab: derived props} Properties derived from the SDSS spectra.} 
	\begin{tabular}{lllll}
	\hline
		Galaxy& 
		E(B-V)[mag]& 
	 	O32& 
		12+log(O/H)&
		N2 index \\
	\hline
	\galone\   &	$0.14\pm0.02$ &	$3.3\pm5.7$ &	$8.04\pm0.02$ &	$0.026\pm0.004$\\ 
	\galtwo\   &	$0.39\pm0.04$ &	$0.28\pm0.09$ &	$8.42\pm0.02$ &	$0.28\pm0.03$\\ 
	\galthree\ &	$0.09\pm0.03$ &	$1\pm0.74$ &	$8.27\pm0.02$ &	$0.17\pm0.017$\\ 
	\galfour\  &	$0.1\pm0.06$ &	$4\pm24$ &	$7.84\pm0.42$ &	$0.003\pm0.016$\\ 
	\galfive\  &	$0.18\pm0.01$ &	$0.9\pm0.2$ &	$8.25\pm0.01$ &	$0.132\pm0.005$\\ 
	\galsix\   &	$0.03\pm0.06$ &	$2.4\pm7.8$ &	$8.11\pm0.04$ &	$0.05\pm0.01$\\ 
	\galseven\ &	$0.13\pm0.04$ &	$1\pm0.8$ &	$8.16\pm0.04$ &	$0.07\pm0.03$\\ 
	\hline
	\end{tabular}
	\end{table}

\subsubsection{Characterizing the observed \lya\ halos}\label{sec: fitting methodology}
	The most common way of characterizing low surface brightness \lya-halo emission is to study azimuthally averaged radial profiles. However, when considering data that has been spatially binned, such as the Voronoi binning used here, doing photometry in simple circular annuli would lead to many cases where the same Voronoi cell contributes to the flux in more than one annulus. The data and errors in adjacent annuli would therefore become correlated, which may be an issue for correctly fitting the surface brightness profiles. 

\begin{figure}
    	\centering
	    \includegraphics[width=0.45\textwidth]{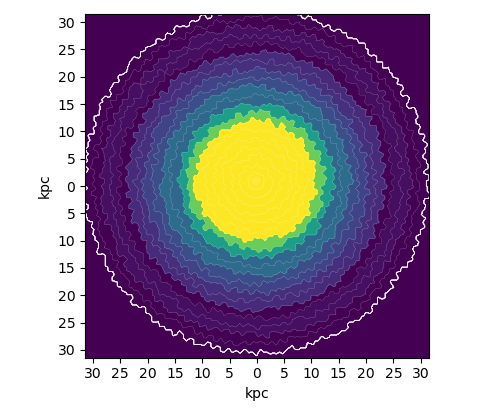}
	    \caption{An illustration of the annular binned F150LP flux of  \galfive.}
	    \label{fig:annulus example}
	\end{figure}

	To overcome this we use a slightly different method, that is based on the Voronoi cells. Each Voronoi cell is assigned a distance to the center of the galaxy that is computed as the mean of the distances to all the pixels that constitute the cell. We create a set of bins that are linear in radius, and then assign Voronoi cells to the bins based on whether their distance lies within the radial bin. This leads to almost circular bins with uneven edges, see Figure\,\ref{fig:annulus example}, but ensures that a pixel is only counted once, even in radial averages. The error on each bin is calculated as the standard deviation of the fluxes of the included Voronoi cells.

	When we construct the radial profiles we also want to limit the  maximum radius to which we fit, in order to avoid numerical biases, and detector edge effects. We therefore adaptively calculate a maximum radius to fit to by stopping the radial profile where two consecutive bins fall under a signal to noise of 0.75. We demanded two consecutive bins to ensure that any continuum absorption in \lya\ did not stop our radial profile. From testing we found that the exact choice of SNR threshold, if taken within reasonable limits (0.5 to 1.5), did not significantly impact the maximum radius selected by the method. The bin sizes are chosen to be small, 0.25 kpc, in the central part of the galaxies (central 1 kpc for all galaxies except \galfive\ where we used 3 kpc due to the complex extended morphology, see Section\,\ref{sample: galfive}) and larger in the outskirts. The larger bin size was chosen for each galaxy to correspond to $\approx10$px, which is slightly larger than the maximum size of a Voronoi cell. This makes sure that the annuli are complete and not broken up. We conducted several tests of binning pattern but found that the results were insensitive to them.

	We then fit the \lya\ as well as the FUV and \halpha\ profiles constructed using the FUV brightest pixel as the center. For FUV and \halpha\ we fit single exponential profiles. For \lya, however, we want to characterize the level of contribution from an extended halo component and to compare the results to high-redshift works. We therefore follow \citet{leclercq2017} and construct the \lya-model as the sum of two exponentials---one core component and one halo component---where the scale length of the core component is taken to be the same as the scale length of the FUV emission, but the amplitude is allowed to vary. The halo component is left unconstrained. 

	Additionally, to characterize the emission of \lya\ properly we need to address the fact that when using filter-based observations, such as our synthesized \lya\ band, continuum absorption can become significant in FUV bright galaxy regions. Therefore we mask all Voronoi tesselations that show negative \lya\ and that lie within 2 kpc of the brightest pixel during the fitting. 

	While this procedure allows us to directly compare results with \citet{leclercq2017}, we can also use the detail in our low-$z$ data to explore other physically motivated models. Almost all our targets (with the possible exception of \galfive, see Section\,\ref{sample: galfive}) show apparent UV emission that is extended compared to the single exponential fit. We therefore extend the previous model by modelling the UV with a 2 component exponential profile instead of a single profile.

\begin{figure}
        \centering
        \includegraphics[width=0.5\textwidth]{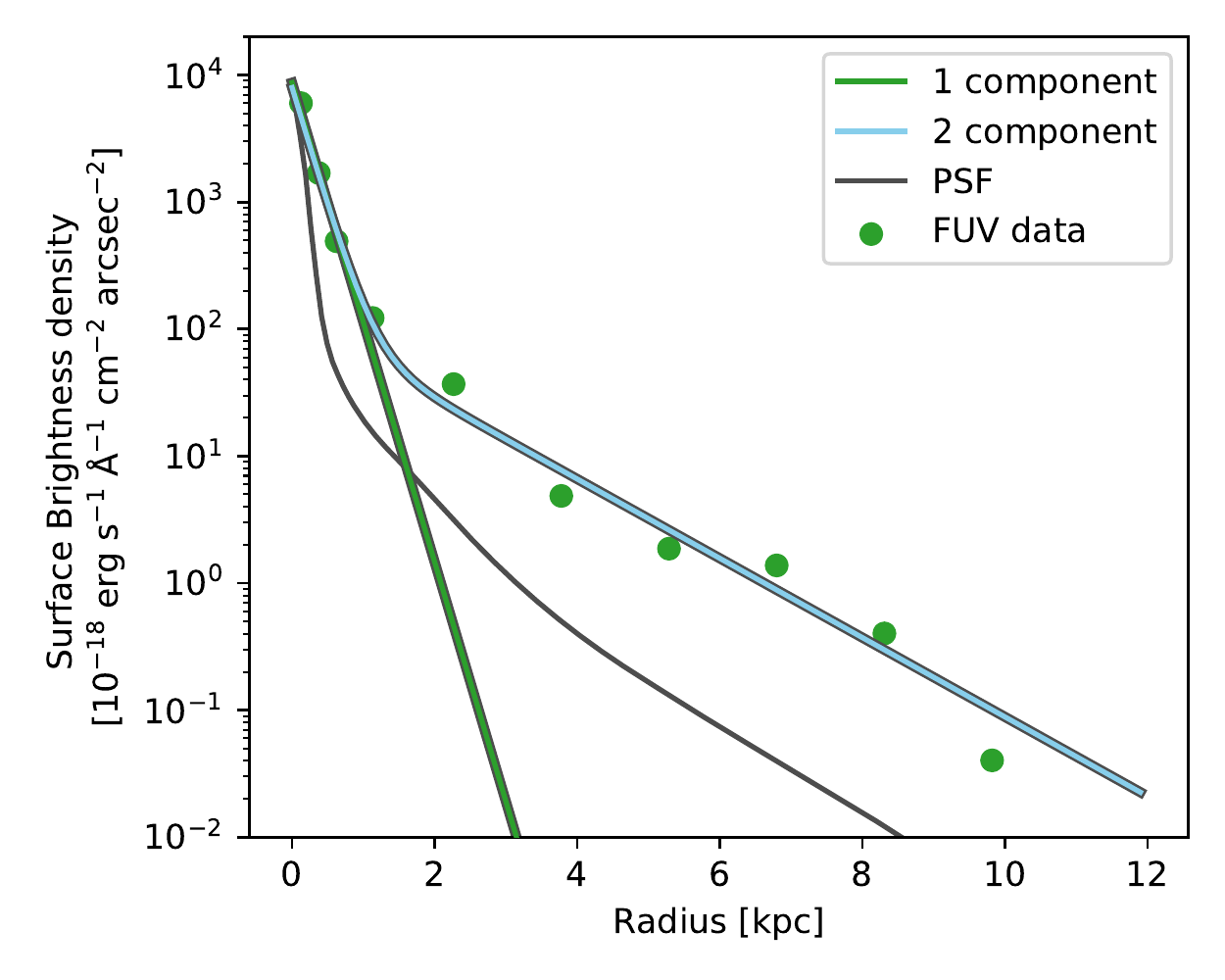}
        \caption{A comparison of the one-component UV model and two-component model for \galone. This quite clearly shows the additional faint extended UV emission that is not captured by a one-component exponential fit. We also show the PSF of the F165LP filter, demonstrating that the extended emission is not due to the wings of the PSF.}
        \label{fig:model comparison}
    \end{figure}

	Figure\,\ref{fig:model comparison} shows the FUV profile of \galone\ and clearly demonstrates this extended low SB component. The 2 component profile, which is shown as the light blue line, follows the points at larger radii far better than the single component model. When comparing the sum of squared residuals outside 3 kpc, we find that the 2 component model reduces this by a factor of more than 2 on average.

	Fitting the high and low SB points well at the same time is challenging however, and we resolved this by fitting the UV data in two steps. First we fit the central profile, as described above. This profile is used to generate a set of constraints for the full two-component fit. Specifically, the central component scale length is constrained to be the same as the one component fit, and the outer component scale length is constrained to be larger than that central one. All other parameters are left free. In this model, we fit \lya\ using three components: one core and one extended component, with scales and relative amplitudes set by the two component FUV fit, and one halo exponential component.

    In Figure\,\ref{fig:model comparison} we also show the PSF of the F165LP filter, noting that the second UV component is significantly brighter than the extended PSF wings even in \galone\ which is one of the most compact sources. We also made an estimate of the potential contribution of nebular continuum to the UV emission based on the \halpha\ surface brightness profile. The conversion factor between \halpha\ emission and nebular continuum is approximately 0.003 $Å^{-1}$ at 1500Å and we find that the estimated nebular continuum is far below the measured UV surface brightness in all our targets. Any dust obscuration would reduce the nebular continuum contribution even further.

	We can further leverage our low-$z$ data by using our \halpha\ observations and by considering the fact that, while the FUV should trace the stellar component that gives rise to the ionizing photons powering the \lya\ emission, \halpha\ should trace the gas where the \lya\ is produced. So we should really be comparing to \halpha\ to determine whether \lya\ is significantly extended and affected by spatial scattering. However, the signal-to-noise of the \halpha\ data is lower than both the FUV and \lya\  and the radius found by the adaptive radial profile (see description above) was quite small for \halpha. Therefore we were only able to use it to constrain the central part of \lya. The approach we took was therefore to fit \lya\ with 2 components, but constraining the center using \halpha\ instead of \lya. The results of this are described in Section\,\ref{sec: Halpha centered fits}.

%% file: results.tex
\section{Results}\label{results}

\subsection{Global \lya\ properties of the sample}\label{sec: global props}

First we place our sample in the context of other \lya-emitting galaxies and contrast global output with other studies.  Global fluxes were measured in circular apertures designed to encapsulate the full \lya\ halo flux, henceforth referred to as global apertures. 
We set the size of the global apertures by iteratively extending them from the UV brightest point and increasing their size by 10 pixels per iteration until the signal-noise-ratio in the next 10px wide annulus was below 1 i.e. it would add more noise than signal. 
The fluxes and luminosities of the \lya, \halpha\ and FUV measured in this aperture are listed in Table\,\ref{tab: Global fluxes and luminosities}, while derived dust extinction, metallicity and O32 $\equiv$ [\oIII]5007$ / $[\oII]3727,29 ratio are shown in Table\,\ref{tab: derived props}. Errors on the \lya\, \halpha\ and FUV fluxes were derived as the standard deviation of the measurements made in each layer of the MC cube (see Section \ref{sec:modelling-the-continuum}).

\begin{figure*}
    \centering
    \includegraphics[width=.9\textwidth]{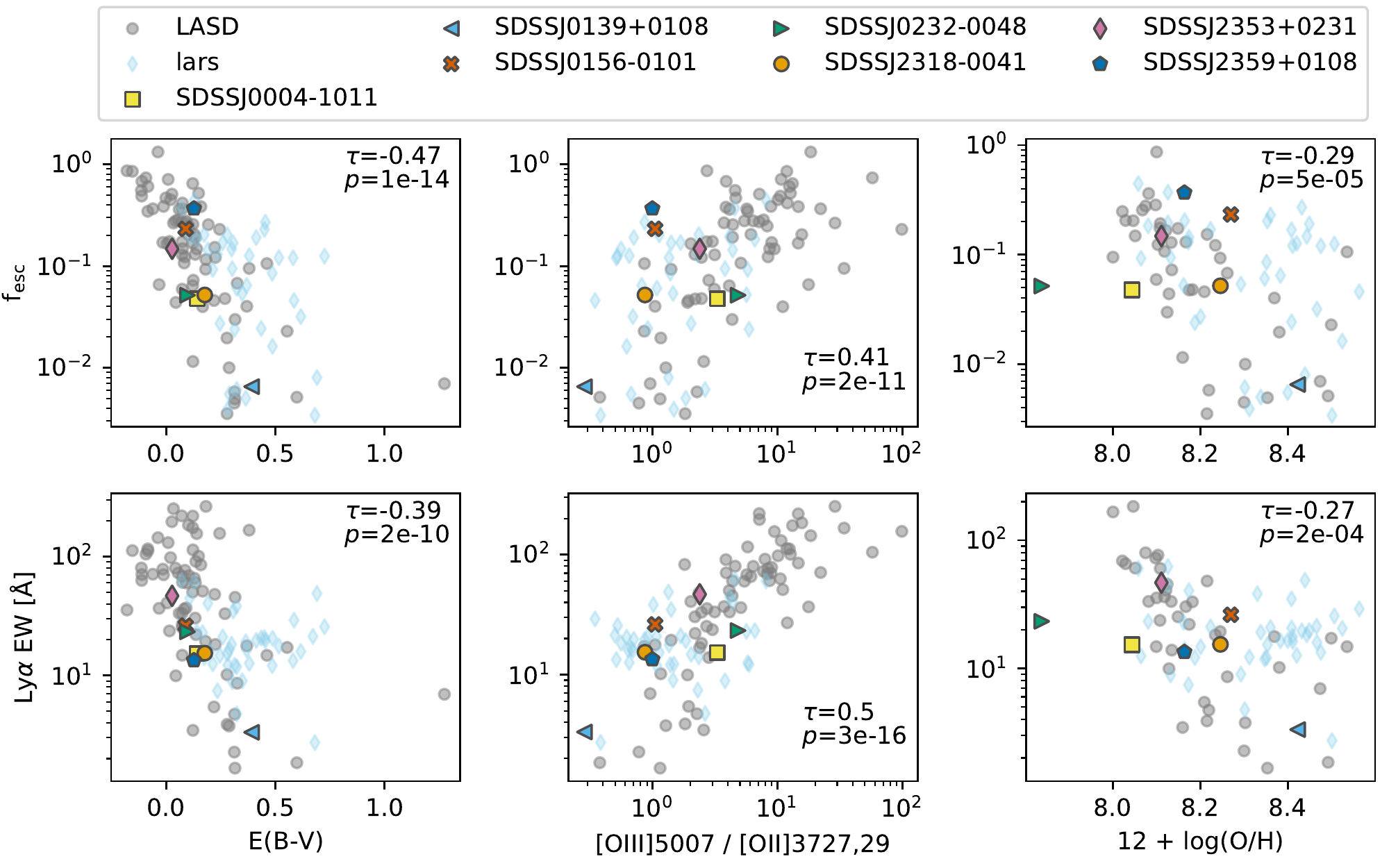}
    \caption{Correlation plots of the global \lya\ properties of our sample. Grey points indicate galaxies with \lya\ measured in COS spectroscopy using the LASD database \citep{runnholm2021}. Light blue diamonds show the distribution of \lya\ properties of the LARS and eLARS galaxies measured in a global aperture. The $\tau$ in each panel indicates the Kendall tau correlation coefficient of all galaxies shown, and $p$ indicates the corresponding p-value.}
    \label{fig:Global correlations}
\end{figure*}

We robustly detect \lya\ ($>5\sigma$) in the global aperture in 6 out of 7 targets. In the remaining target the global aperture shows a $\sim 3 \sigma$ detection. We stress that this is in a large aperture and therefore is heavily impacted by central absorption. We detect \lya\ emission around all galaxies when considering binned annuli. The escape fractions and equivalent widths of \lya\ are presented in Table\,\ref{tab: fesc}. The \lya\ escape fraction is defined as 
\begin{equation}
f_{\mathrm{esc}} = \frac{\mathrm{Ly}\alpha}{8.7 \times \mathrm{H}\alpha}
\end{equation}
where \halpha\ is corrected for dust extinction and 8.7 is the assumed intrinsic \lya\ to \halpha\ ratio. We find that our sample spans a large range in escape fractions, from 0.7\% to 37\%. The equivalent widths ranges from 4.4 \AA\ in \galtwo\ to 57 \AA\ in \galsix. 

In Figure\,\ref{fig:Global correlations} we show the correlations of the \lya\ escape fraction and EW  with some nebular properties of the gas. Specifically we show correlations with dust extinction (E(B-V)), ionization parameter traced by the O32 ratio, and the metallicity of the galaxies measured from the O3N2 index using the \citet{marino2013} calibration. In addition to the seven galaxies presented in this work, we also show data for the LARS sample from \citet[][redshift range $0.028 \leq z \leq 0.181$]{Melinder2023} and a large collection of low redshift ($0.02 \leq z \leq 0.44$) COS observations taken from the Lyman Alpha Spectral Database \citep[LASD,][and references therein]{runnholm2021, Hayes2023}. The optical data for all galaxies comes from the SDSS.  We note that these different samples have very different selection effects and observational methods, and would not necessarily be comparable; we show this comparison primarily to illustrate that our sample does not appear different from any of the others, regardless of how the measurements were made. 
We see substantial scatter in all of the displayed relations, but when taken in conjunction with large archival samples some clear trends do emerge. In fact, as indicated by the Kendall-tau correlation metrics displayed in the panels, all of the correlations become statistically significant with some  correlations, such as \lya\ EW and O32 ratio  and the anticorrelation between \lya\ f$_\mathrm{esc}$ with dust extinction (E(B-V)) having p-values as low as $10^{-16}$ and $\tau\approx0.5$.

\begin{table*}
\caption{\label{tab: Global fluxes and luminosities} Fluxes and luminosities in \lya, \halpha, and FUV.} 
\begin{tabular}{lllllll} 
\hline
	Galaxy& 
	Ly$\alpha$ flux $^{1}$& 
	Ly$\alpha$ luminosity$^{2}$& 
	H$\alpha$ flux $^{1}$& 
	H$\alpha$ luminosity$^{2}$& 
	FUV flux $^{3}$& 
	FUV luminosity$^{4}$\\
\hline
\galone\   &	$560\pm19$ &	$96\pm3$ &	$970\pm140$ &	$166\pm24$ &	$29.6\pm0.2$ &	$5.063\pm0.036$\\
\galtwo\   &	$42\pm16$ &	$13\pm5$ &	$298\pm31$ &	$91\pm9$ &	$9.7\pm0.13$ &	$2.95\pm0.04$\\ 
\galthree\ &	$1220\pm51$ &	$276\pm12$ &	$490\pm158$ &	$111\pm36$ &	$36.7\pm0.5$ &	$8.29\pm0.12$\\ 
\galfour\  &	$75\pm10$ &	$23\pm3$ &	$130\pm55$ &	$41\pm17$ &	$2.46\pm0.08$ &	$0.762\pm0.024$\\ 
\galfive\  &	$1433\pm40$ &	$276\pm8$ &	$2100\pm130$ &	$405\pm26$ &	$74.3\pm0.43$ &	$14.30\pm0.08$\\ 
\galsix\   &	$556\pm15$ &	$90\pm2.4$ &	$405\pm60$ &	$66\pm10$ &	$9.645\pm0.169$ &	$1.567\pm0.027$\\ 
\galseven\ &	$720\pm30$ &	$150\pm6$ &	$170\pm100$ &	$35\pm21$ &	$42.5\pm0.3$ &	$8.88\pm0.06$\\ 
\hline
\end{tabular}

\hbox{$^{1}$Flux in units of $10^{-17}$~erg~s$^{-1}$~cm~$^{-2}$. }

\hbox{$^{2}$Luminosity in units of $10^{40}$ erg~s$^{-1}$.}

\hbox{$^{3}$Flux density in units of $10^{-17}$~erg~s$^{-1}$~cm~$^{-2}$~Å$^{-1}$. }

\hbox{$^{4}$Luminosity density in units of $10^{40}$ erg~s$^{-1}$~Å$^{-1}$.}
\end{table*}

\begin{table}
\caption{\label{tab: fesc} Table of escape fractions, equivalent widths of \lya, measured halo fractions from the 3 component model and the size of the global aperture used. } 
\begin{tabular}{lllll}
\hline
	Galaxy& 
	Ly$\alpha$ f$_\mathrm{esc}$& 
	Ly$\alpha$ EW&
  HF&
	Aperture \\
	& 
	& 
	[Å]& 
  &
	[arcsec]\\
\hline
\galone\   &	$0.048\pm0.007$ &	$15.3\pm0.4$ & $0.68\pm0.02$ &	$3.6$\\ 
\galtwo\   &	$0.007\pm0.003$ &	$3.3\pm0.9$& $0.30\pm0.30$ &	$1.2$\\ 
\galthree\ &	$0.23\pm0.08$ &	$26.2\pm0.9$& $0.78\pm0.01$ &	$4.0$\\ 
\galfour\  &	$0.05\pm0.02$ &	$23.3\pm2.5$& $0.58\pm0.09$ &	$1.2$\\ 
\galfive\  &	$0.052\pm0.004$ &	$15.4\pm0.4$& $0.90\pm0.17$ &	$4.8$\\ 
\galsix\   &	$0.15\pm0.02$ &	$46.7\pm1.2$& $0.81\pm0.08$ &	$1.6$\\ 
\galseven\ &	$0.370\pm0.22$ &	$13.4\pm0.4$& $0.48\pm0.05$ &	$2.8$\\ 
\hline
\end{tabular}
\end{table}

\import{./}{sample_discussion}

\subsection{Ly$\alpha$ halo sizes}\label{sec: halosizes}

\begin{figure}
    \centering
    \includegraphics[width=0.5\textwidth]{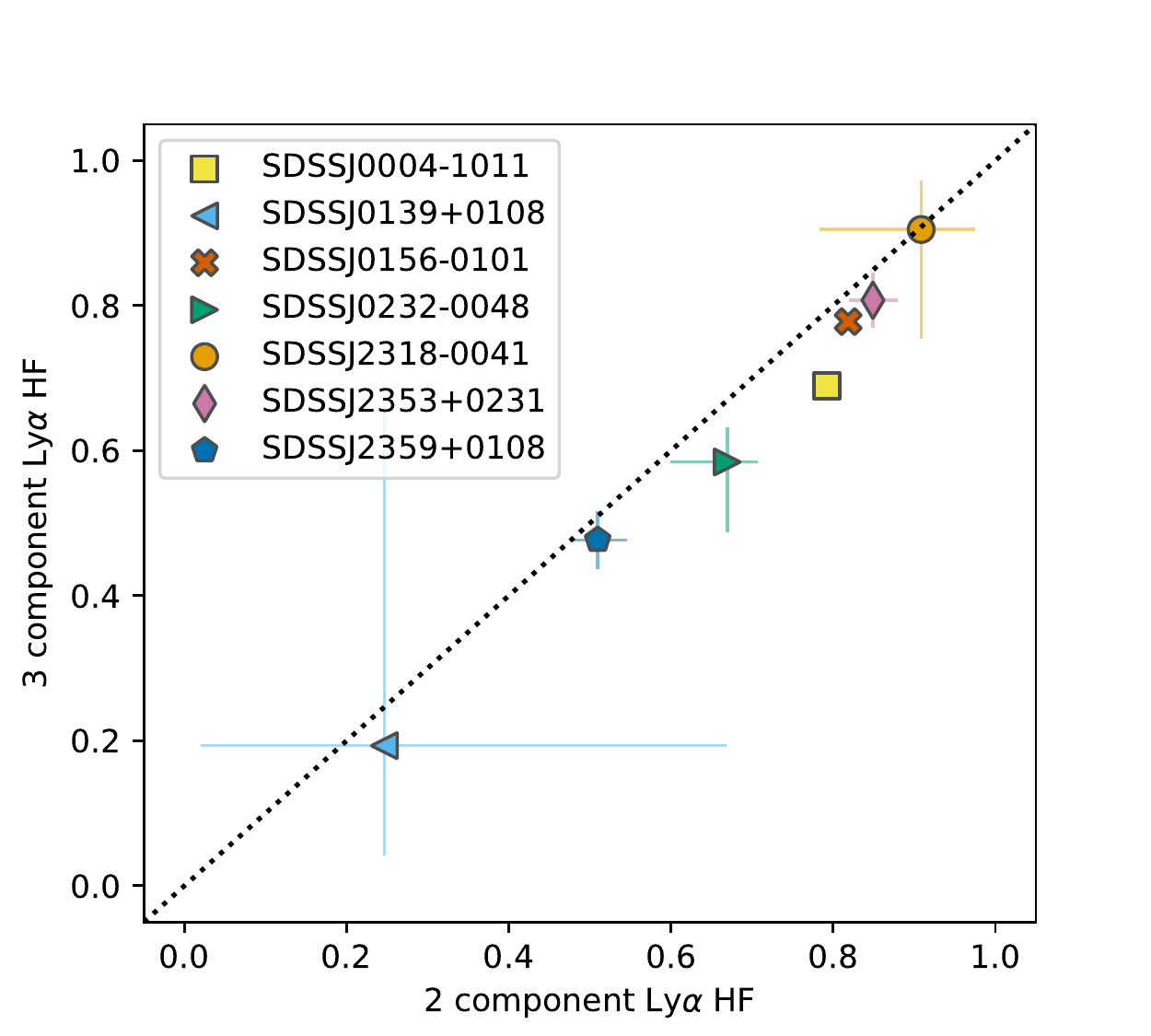}
    \caption{Comparison of the \lya\ halo fractions derived using a 3 component \lya\ model (y-axis) to a 2 component \lya\ model (x-axis). }
    \label{fig:3vs2_halofractions}
\end{figure}

In this Section we discuss the comparison between out targets and high redshift observations. We will primarily consider the results from the 2 component model here as that is most directly comparable to the work of \citet{leclercq2017}, in particular when comparing UV and \lya\ scale lengths. It is worth noting that  the comparison of halo fractions would be largely unaffected by using the 3-component model, as demonstrated by Figure\,\ref{fig:3vs2_halofractions} which shows the halo fractions of the 2 component model versus the halo fractions of the 3 component model.
We find that the new model with 2 FUV components, can describe the full FUV profile significantly better, but that the extended UV is very faint compared to the peak core flux. In particular, the relative core to halo flux level in the UV is much higher than in the \lya, which means that the additional UV component has only marginal effects on the \lya\ halo fit. The result is that the halo fractions that we measure are largely unchanged, with decreases on the order of a few per cent, as shown in Figure\,\ref{fig:3vs2_halofractions}.

\begin{figure*}
  \centering
  \includegraphics[width=\textwidth]{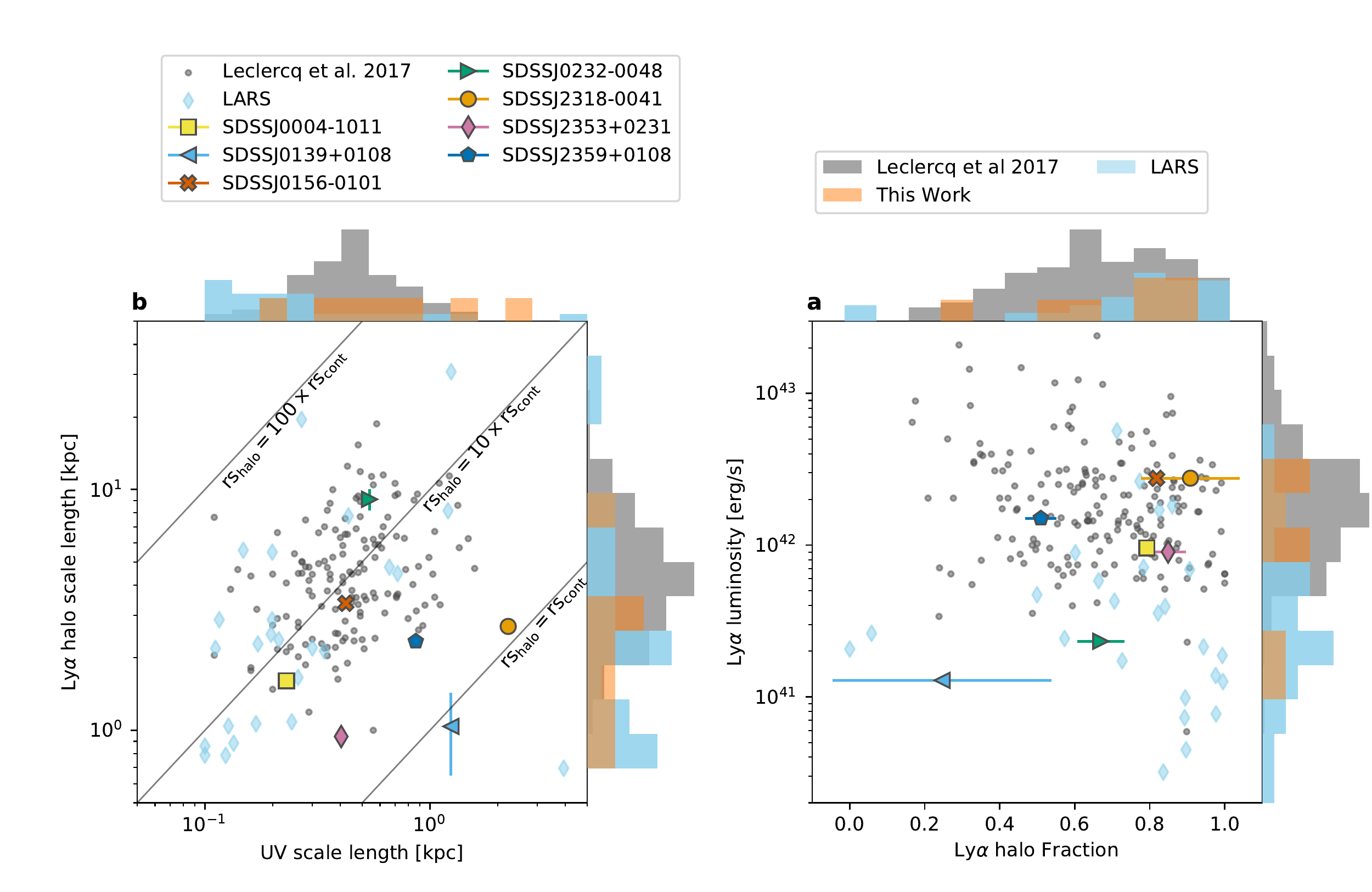}
  \caption{Panel \textbf{a}: Relation between halo scale length and UV scale length. Gray points indicate galaxies from \citet{leclercq2017}. Light blue points are measurements from the LARS sample \citet{rasekh2021}. Diagonal lines denote lines of constant relation between the \lya\ and FUV scale lengths, denoted $rs_\mathrm{halo}$ and rs$_\mathrm{cont}$ respectively.  
  Panel \textbf{b}: Distribution of measured halo fractions versus the \lya\ luminosity of the galaxy. Histograms show the distributions of the points for the three samples (MUSE, LARS and this work). }
  \label{fig:Muse standard comparison}
\end{figure*}

Panel \textbf{a} of Figure\,\ref{fig:Muse standard comparison} shows the scale length of the \lya\ halo component versus the scale length of the UV emission. The grey points from \citet{leclercq2017} are scattered around the 10:1 line and the majority of our galaxies fall within this distribution. However, our galaxies tend to lie toward smaller \lya\ over FUV scale length ratios, with the smallest ratios being those of \galfive\ and \galtwo. One should  keep in mind that \galtwo\ has a very weak \lya\ detection and thus no extended \lya\ halo is present in our data, and that \galfive\ has a very complex morphology which leads to a single exponential decline in the light profile rather than a clear core and halo as seen in other targets. To establish whether our seven targets are consistent with the high-z distribution we performed a  two sample KS test on the \lya\ halo to FUV scale length ratios. We found that our sample has a statistically significantly smaller ratio (mean ratio 5.6 kpc vs 12 kpc in \citet{leclercq2017}), with a p-value of 0.01. When \galtwo\ and \galfive\ are excluded the mean is 7.4 kpc and the statistical significance drops to 0.26, i.e. the distributions are not significantly different, although it should be noted that the number of datapoints is in this case very low.

We note that the LARS points have a greater overlap with high-redshift points in the FUV to \lya\ ratio, but also that they extend to smaller sizes both in FUV and \lya. This is most likely not due to LARS galaxies being smaller than galaxies observed by \citet{leclercq2017}, but rather that the observations are resolving internal structures (star-forming clumps) in these low-$z$ galaxies. At the most compact end we are observing scale lengths as small as 0.2 kpc, which would correspond to 0.02 arcsec, or about half a HST pixel at $z$=3, which would not be observable.  Nevertheless, if we include the LARS points in the size-ratio analysis the difference between the low and high-redshift distribution is no longer significant (p-value of 0.14).

Panel \textbf{b} of Figure\,\ref{fig:Muse standard comparison} shows the \lya\ halo luminosity as measured from the fitted model against the halo fraction (HF) of \lya. We define the halo fraction as the fraction of model flux coming from the halo component compared to the total model flux when both are integrated to infinity. It is clear that the distribution of our galaxies is quite consistent with the high-redshift results which we confirm using a 2 sample KS test (p-value=0.6). Our seven targets mostly scatter between 50 and 85 \% contribution from the halo with one target having a lower HF of 0.25 but with very significant measurement uncertainties. This panel also shows the distribution of points from the LARS sample, and when those are included we see that the relative halo contributions at low redshifts appears to be higher than at high redshifts, although it should be noted that this is largely driven by points that lie at \lya\ luminosities below what is observed in the \lya\ selected sample of \citet{leclercq2017}.

%% file: sample_discussion.tex
\subsection{Detailed discussion of the sample}\label{sec: sample discussion}
In this Section we discuss our observations and our sample of galaxies in more detail since some of our objects show interesting features in all observed bandpasses and careful  consideration of the surface brightness profiles and fits are useful for interpreting subsequent results. 

For each target we present one figure akin to Figure \ref{fig:SDSSJ0004}. Panel \textbf{a} shows a three-color image of \halpha\ in red, FUV in green and \lya\ in blue with arcsinh scaling. The rest of the top row (Panels \textbf{b}--\textbf{d}) shows the individual bands with the same color assignment but log scaled. The second row (Panels \textbf{e}--\textbf{g}) show binned radial profiles and multiple component fits to the data (2 components for UV and \halpha\ and 3 components for \lya).

\subsubsection{\galone}
\begin{figure*}
  \centering
  \includegraphics[width=\textwidth]{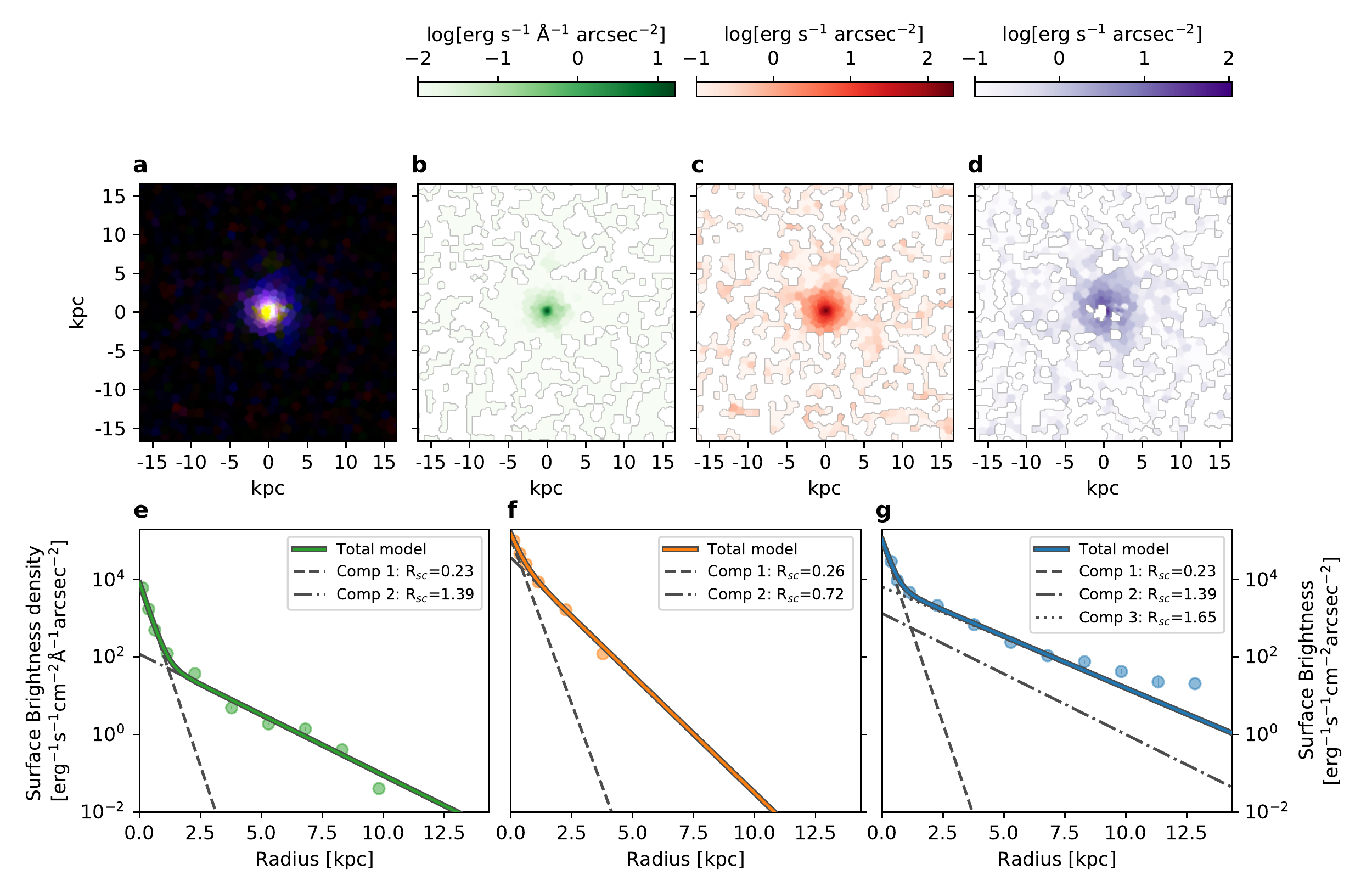}
  \caption{Images and surface brightness profiles for \galone. All surface brightnesses are given in units of $10^{-18}$ erg$^{-1}$ s$^{-1}$ Å$^{-1}$arcsec$^{-2}$:
  \textbf{a)} \lya, FUV, and \halpha\ composite image with arcsinh scaling to highlight low surface brightness. 
  \textbf{b)} Log-scaled F165LP UV image;
  \textbf{c)} Log-scaled \halpha\ image; 
  \textbf{d)} Log-scaled \lya\ image;
  \textbf{e)} binned radial surface brightness profile of the FUV, together with the best fit 2 component model of the emission. Note that the FUV emission is given in flux densities rather than fluxes, i.e. [erg s$^{-1}$~\AA$^{-1}$~cm$^{-2}$~arcsec$^{-2}$] since it is continuum emission rather than integrated line emission;
  \textbf{f)} binned radial surface brightness profile of the \halpha, together with the best fit 2 component model of the emission;
  \textbf{g)} binned radial surface brightness profile of the \lya, together with the best fit 3 component model of the emission;
  }
  \label{fig:SDSSJ0004} 
\end{figure*}

  \galone\ is shown in Figure\,\ref{fig:SDSSJ0004} and has a notably compact and circular appearance even in high spatial resolution HST imaging. 
  The total escape fraction of \lya\ is quite low, at 5\%, due to strong continuum absorption in the center of the galaxy. The UV shows a sharp peak with a scale length of 0.23 kpc and then a clear flatter outer component (scale length 1.39 kpc) which is much fainter (the amplitude difference is approximately a factor of 100). Despite taking the extended UV emission into account we find that there is still evidence for extended halo emission in  the \lya-profile with the extended outer component showing a distinctly higher amplitude relative to the center than in the FUV as well as a slightly flatter slope (scale length of 1.65 kpc). There are also hints that the \lya\ is flattening out even further at radii above 10 kpc however the remaining data are too low signal-to-noise to determine this with any certainty. The measured \lya\ halo fraction of this galaxy is $0.69\pm0.03$. 

  Turning to the \halpha\ surface brightness profile, we note firstly that it is only well traced out to a radius of $\sim4$ kpc. The data seem well described by two exponential profiles. The slope of the \halpha\ is distinctly shallower than the core component of the UV. This could indicate that ionizing photons produced in the most UV bright regions travel some distance before being absorbed and ionizing the gas. We will discuss this further in Section\,\ref{sec: Halpha centered fits}.

\subsubsection{\galtwo}
\begin{figure*}
  \centering
  \includegraphics[width=\textwidth]{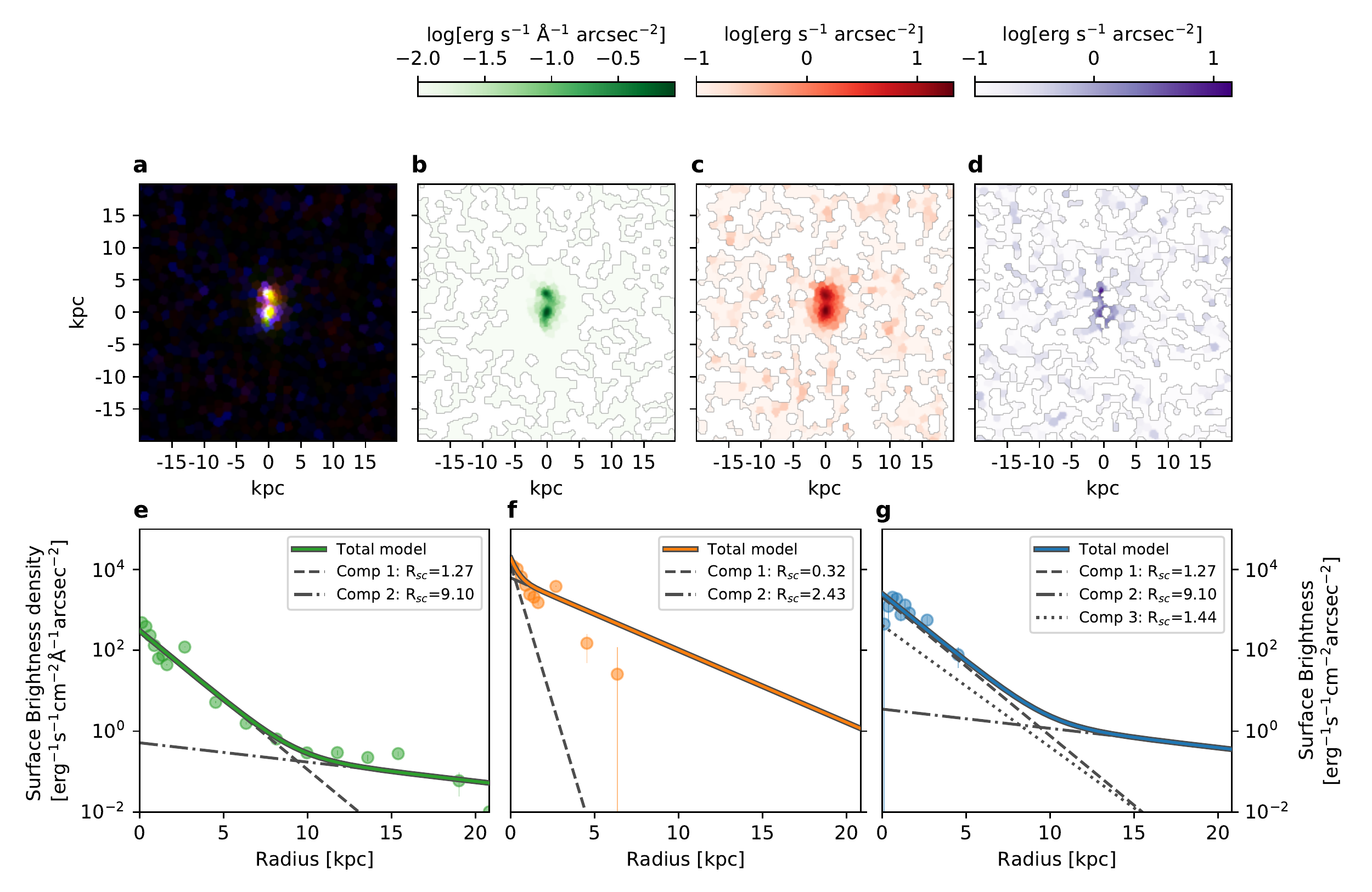}
  \caption{\lya\ \halpha\ and FUV profiles for \galtwo. For a detailed description of each Panel see Figure\,\ref{fig:SDSSJ0004}}
  \label{fig:SDSSJ0139}
\end{figure*}

The morphology of \galtwo\ is dominated by two star forming clumps which can also be seen as two distinct bumps in  the radial profile. Despite the clumped morphology, the FUV radial profile is reasonably well described by a central exponential decline and there is compelling evidence for extended FUV emission (Panel \textbf{e} of Figure\,\ref{fig:SDSSJ0139}) that is well captured by a two-component FUV model. The galaxy shows strong central absorption but has weak Lya emission around the clumps. Panel \textbf{g} demonstrates that the \lya\ is very weak in this target and that constraining the contribution from a \lya\ halo in this case is difficult. This is reflected in the uncertainty on the measured halo fraction $0.19\pm0.28$, which is consistent with zero halo contribution, as we would expect from looking at the few \lya\ datapoints. 

The \halpha\ emission is quite weak outside the very central parts of the galaxy, and the pronounced two clump morphology of the system drives the fit to find a flat second component. One could consider several ways of treating this morphology, including removing the datapoint at 3 kpc which would return the profile to an approximate exponential decline, or modelling this extra component using  an additional core component modelled in two dimensions. However, doing detailed 2-D modelling of our targets is beyond the scope of this work. We will, however, further examine the \halpha\ properties of these galaxies in a forthcoming paper using deep ground-based observations.

\subsubsection{\galthree} 
\begin{figure*}
  \centering
  \includegraphics[width=\textwidth]{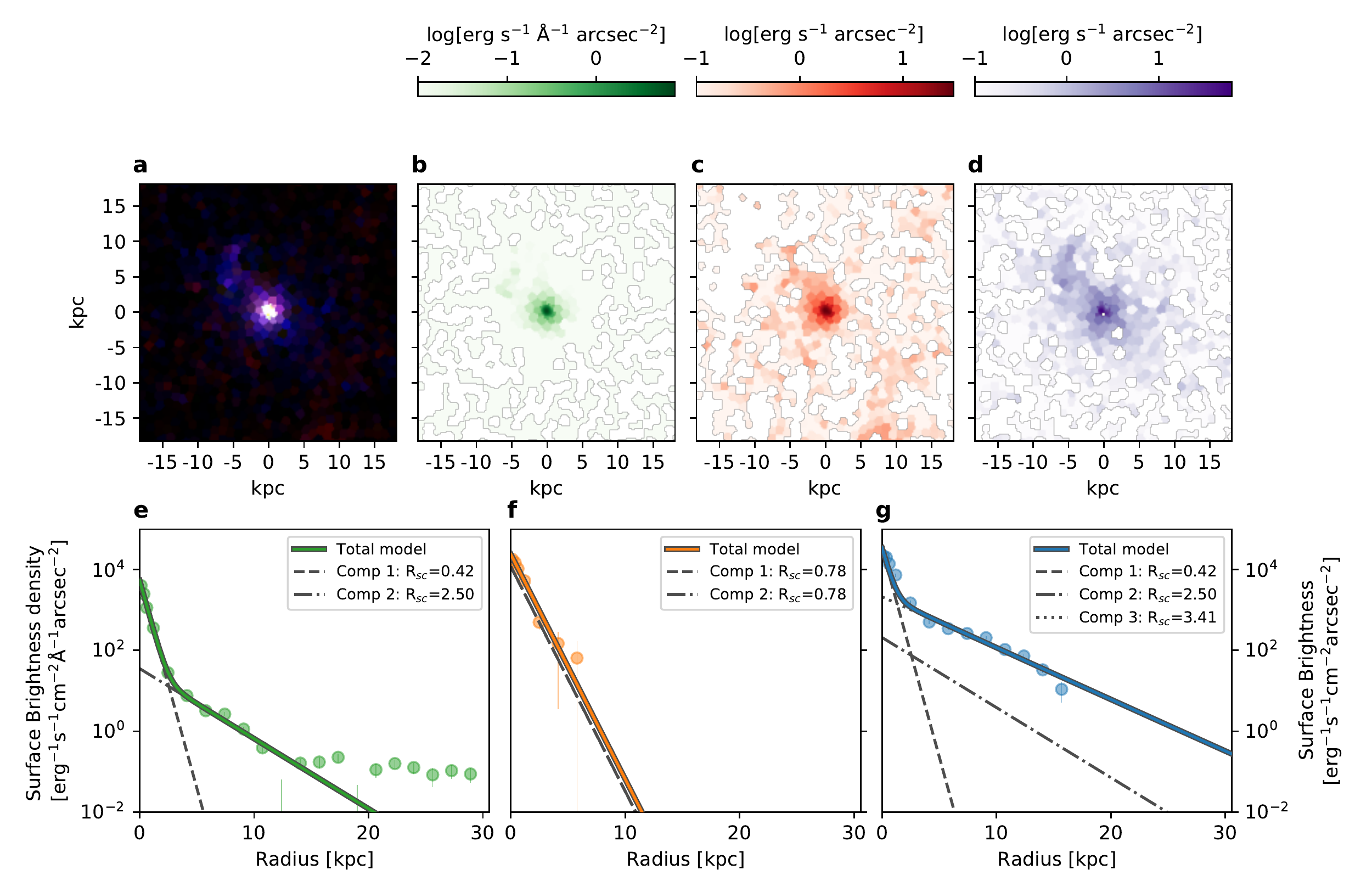}
  \caption{\lya\ \halpha\ and FUV profiles for \galthree. For a detailed description of each Panel see Figure\,\ref{fig:SDSSJ0004}}
  \label{fig:SDSSJ0156}
\end{figure*}

\galthree\ (Figure\,\ref{fig:SDSSJ0156}) is morphologically similar to \galone\ i.e. circular and compact. However, as is clear from the three color image there is also evidence for a secondary structure to the north east of the main galaxy, which could be a fainter star forming knot or a remnant of an interaction that possibly caused the central starburst. The \lya\ emission from \galthree\ is significantly stronger than \galone (with a total luminosity of $1.2\times10^{42}$ erg s$^{-1}$), and it has a global escape fraction of 23\%. One possible reason for this is lower central continuum absorption. While we cannot measure the level of continuum absorption with our data we note that this galaxy shows fewer absorption dominated pixels, i.e. pixels that are negative in the \lya\ image, in the UV bright regions than the other targets and the implied escape fraction for the central 1 kpc is around 15\%. 

Examining the surface brightness profiles, beginning with the FUV in Panel \textbf{e}, we note several similarities with \galone, specifically a steep bright central core component and a flat outer component. We do note however that in this case there are a significant number of UV points that show an even flatter slope outside of 20 kpc before the signal to noise drops below our threshold.
Similar extended emission is also seen in the \lya\ SB profile, but the \lya\ halo component is flatter and significantly brighter relative to the core brightness than its FUV counterpart, leading to a large measured halo fraction of $0.78\pm0.012$

\subsubsection{\galfour} 
\begin{figure*}
  \centering
  \includegraphics[width=\textwidth]{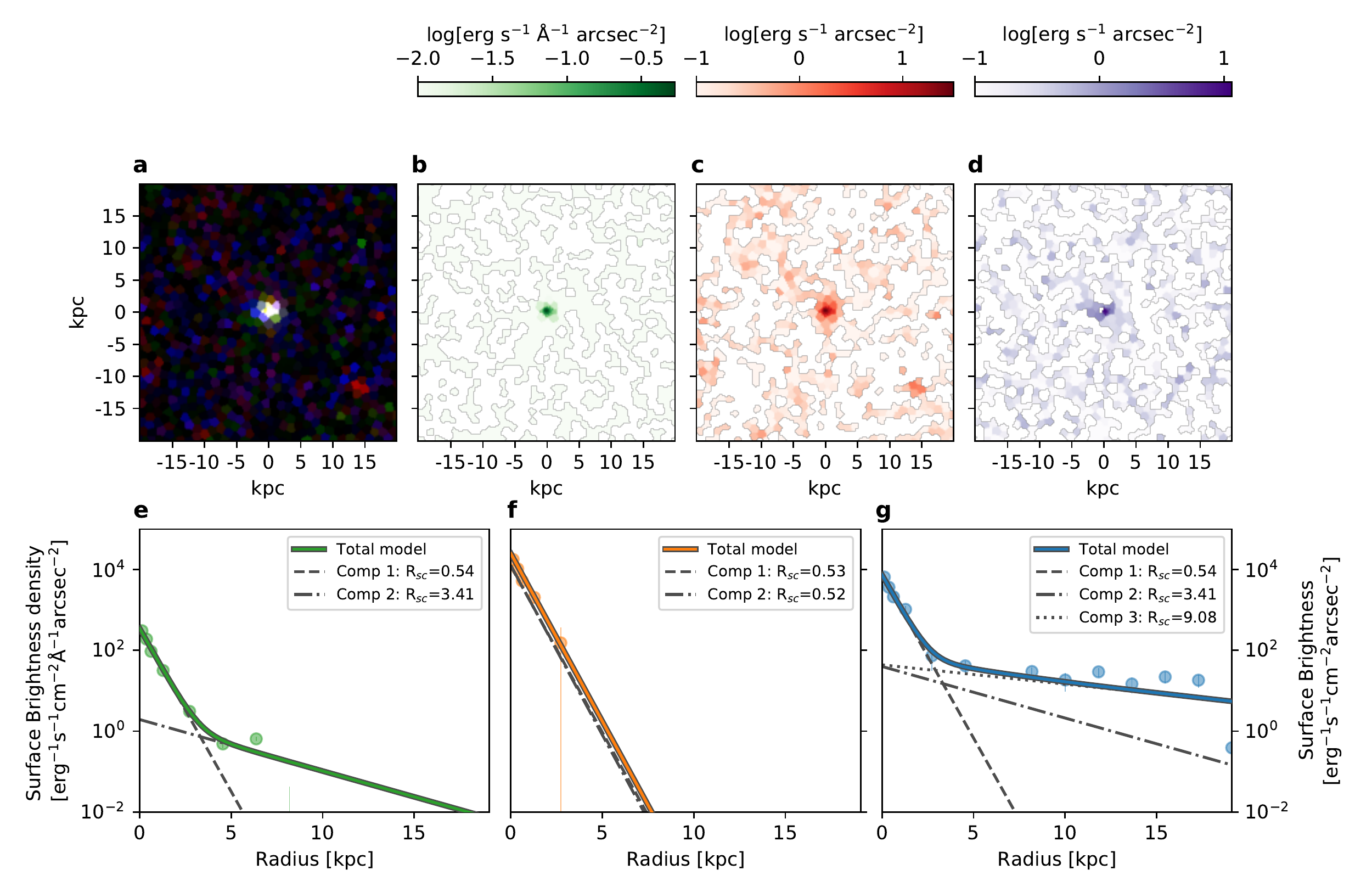}
  \caption{\lya\ \halpha\ and FUV profiles for \galfour. For a detailed description of each Panel see figure\ref{fig:SDSSJ0004}}
  \label{fig:SDSSJ0232}
\end{figure*}

\galfour\ (Figure\,\ref{fig:SDSSJ0232}) is the smallest galaxy in our sample in both angular and physical size. The FUV profile shows very compact emission and some hints of a turnover at a radius of 5 kpc. The \halpha\ shows similar behaviour but unfortunately cannot be traced out to this radius leaving us unable to confirm the presence of a similar turnover. The \lya\ profile has an inner core that is well described by the FUV and \halpha\ radial profile fits, but has a distinct and very extended secondary emission component that is significantly flatter than the FUV secondary component. However, the strength of this extended emission relative to the core is lower than in \galone\ and \galthree, leading to a halo fraction of $0.58\pm0.09$.

\subsubsection{\galfive}\label{sample: galfive} 
\galfive\ is the largest and structurally most complicated of the galaxies in our sample. The main structure of the galaxy in the FUV is composed of a number of star-forming clumps, many of which also emit bright H$\alpha$. There are also additional structures visible both in FUV and \halpha, including a series of emission spots tracing a faint arm-like feature to the north east, and a strong clump of \halpha\ emission to the north west. 

Attending now to the radial profile of the FUV in Panel\,\textbf{d}, we see that it is quite uneven and `bumpy', reflecting the clumpy structure of the galaxy. However, unlike the profiles of the other targets, it follows a relatively straight exponential decline, as opposed to a bright central emission core and a faint extended component. The FUV points do show a slight flattening at the very highest radii, around 20 kpc, but the SNR of the data is insufficient to trace this further. This smooth decline in surface brightness leads to issues with our model description. 
The data itself strongly resembles a single exponential, and this is corroborated by the fit returning similar scale lengths for both components but one having a smaller amplitude.
Our conclusion, in this case, is that a single exponential decline is the best description of the surface brightness profile while stopping short of spatially decomposing the galaxy and simultaneously fitting each individual clump.

Turning now to the \halpha\ and \lya\ profiles we note that they are very similar to the UV, both following smooth exponential declines. However, we do note that the scale length of the \halpha\ profile is slightly larger than that of the FUV, and that the \lya\ scale length is in turn slightly larger than that of \halpha. Together this indicates that ionizing photons travel some distance before ionizing hydrogen atoms, creating \halpha\ and \lya\ photons and potentially that \lya\ photons then scatter even further from the site of creation. One should again keep in mind that we cannot trace \halpha\ as far as \lya\ and that the slope of the \lya\ may be impacted by those large radius points. The fact that a single exponential dominates the fit leads to a large halo fraction of $0.90\pm0.17$.

\begin{figure*}
  \centering
  \includegraphics[width=\textwidth]{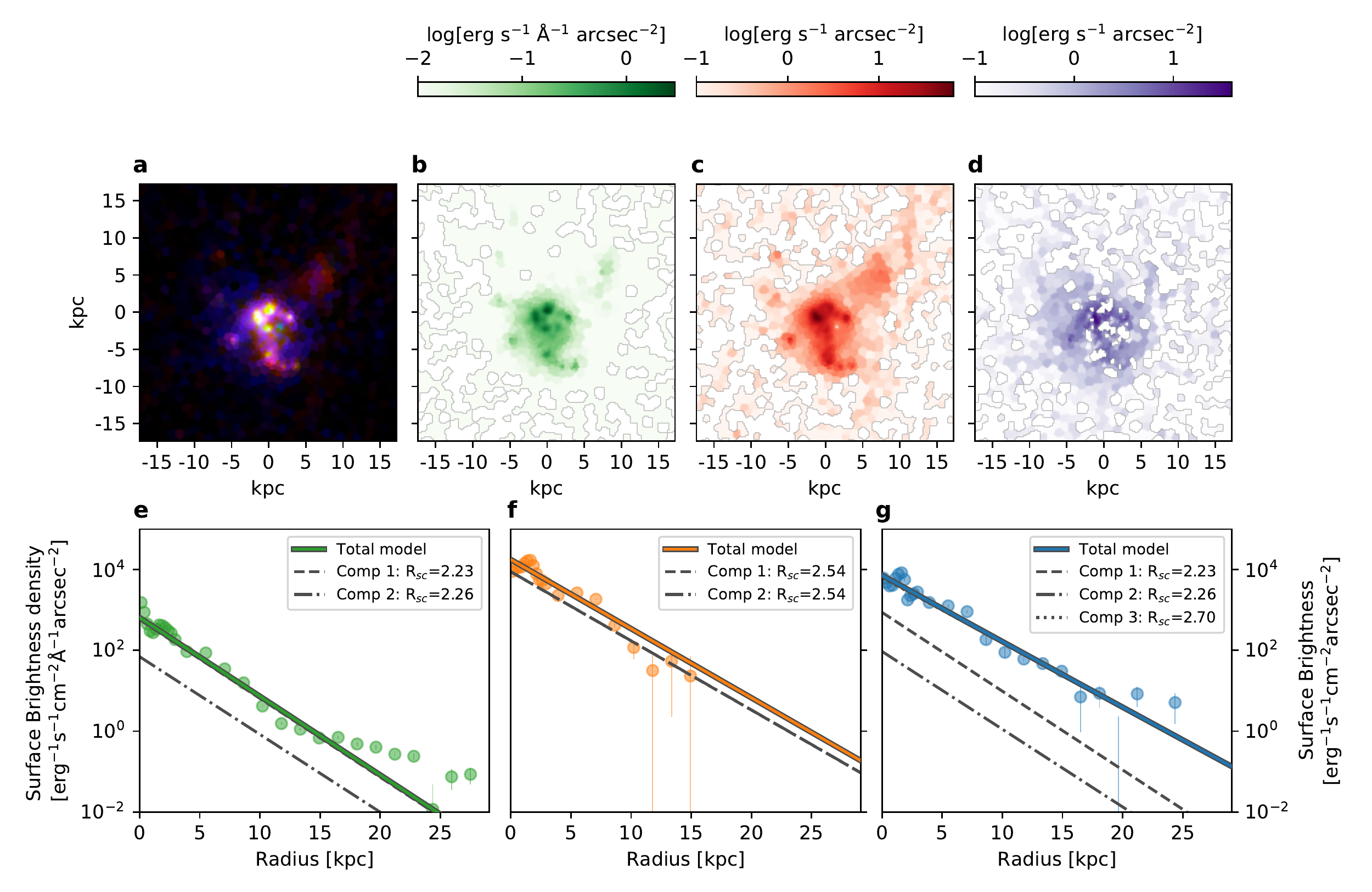}
  \caption{\lya\ \halpha\ and FUV profiles for \galfive. For a detailed description of each Panel see Figure\,\ref{fig:SDSSJ0004}}
  \label{fig:SDSSJ2318}
\end{figure*}

\subsubsection{\galsix} 
\begin{figure*}
  \centering
  \includegraphics[width=\textwidth]{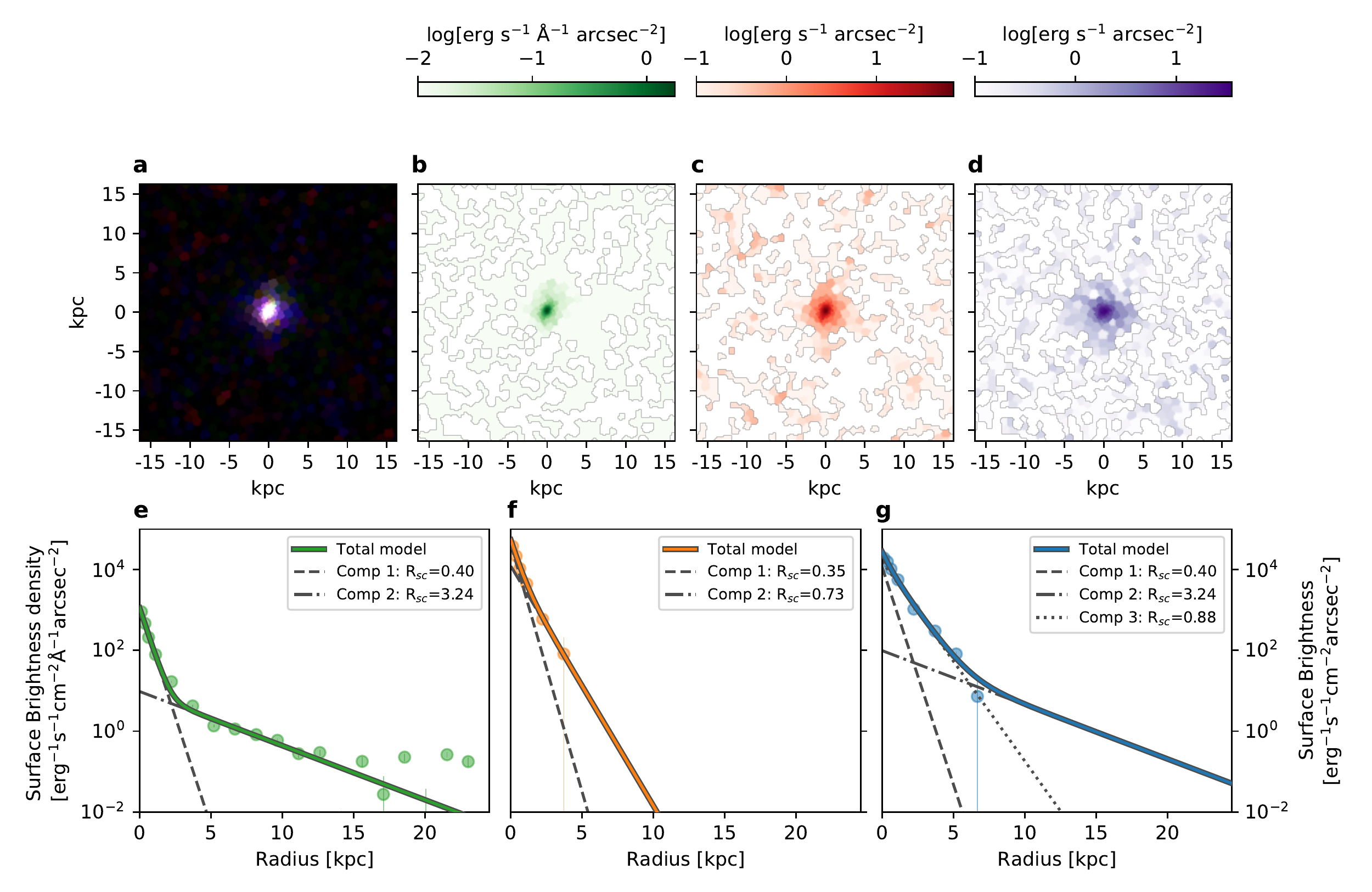}
  \caption{\lya\ \halpha\ and FUV profiles for \galsix. For a detailed description of each Panel see Figure\, \ref{fig:SDSSJ0004}}
  \label{fig:SDSSJ2353}
\end{figure*}

While the overall morphology of \galsix\ is similar to \galone\ and \galthree, Panel\,\textbf{a} of Figure\,\ref{fig:SDSSJ2353} and Figure\,\ref{fig:optical images} show that it is more extended along one axis. Contrasting Panels \textbf{a} and \textbf{c}, we also note that the \lya\ appears significantly more circular than the FUV. Similar contrasts between UV and \lya\ have been observed in edge on systems, such as Mrk 1486 \citep{duval2016a, rasekh2021}, and may be due to preferential \lya\ escape in the direction of galactic outflows emanating from the plane of the galaxy.

Attending next to the surface brightness profiles, we see that the FUV has a very distinct core with a scale length of 0.4 kpc and a secondary faint component with a scale length of $\approx3.2$ kpc. There are also hints at a further, even flatter component at very large radii. We are unable to trace the \halpha\ profile very far but see that it is well parametrized with 2 components, with a sharp central peak and then a flatter decline, but the number of data points is low and the profile can be adequately described with a single profile with a scale length around 0.7 kpc, i.e. larger than the core scale length of the UV.

The \lya\ shows some interesting behaviour, with a distinctly flatter decline than the FUV profile, almost directly from the center. This, taken in conjunction with the \halpha, may indicate that ionizing photons are traveling some distance from the center before ionizing. The data has insufficient signal to noise to trace the \lya\ far enough to tell whether the secondary FUV emission also manifests in the \lya. However, the non-detection of \lya\ at large radii may indicate that the secondary FUV component is coming from a stellar population that does not produce significant ionizing radiation, and hence, no \lya. This situation does not match a classical extended halo. However, if we interpret the core FUV as the primary ionizing component, the \lya\ extension essentially describes a halo around this component. We will refer to this as an inner excess to differentiate it from a more classical halo. The fraction of emission from this inner excess is quite large: $0.81\pm0.08$. 

\subsubsection{\galseven} \label{sec: details J2359}
\begin{figure*}
  \centering
  \includegraphics[width=\textwidth]{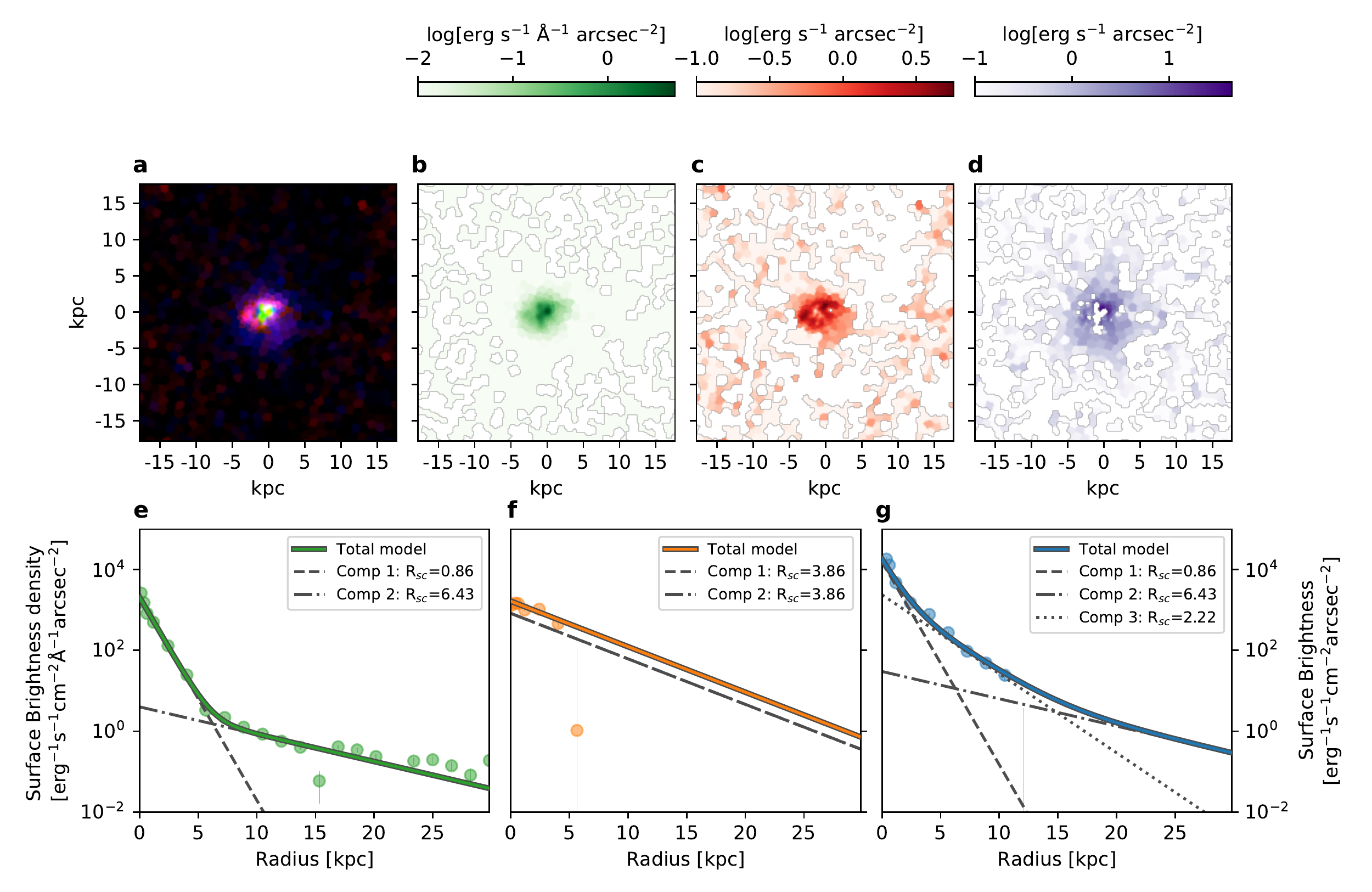}
  \caption{\lya\ \halpha\ and FUV profiles for \galseven. For a detailed description of each Panel see figure \ref{fig:SDSSJ0004}.}
  \label{fig:SDSSJ2359}
\end{figure*}

\galseven\ is perhaps one of the most interesting objects in our sample. In the optical image (last Panel of Figure\,\ref{fig:optical images}) the galaxy does not appear particularly different from the rest of the sample, with a compact morphology closely resembling that of \galfour\ and \galsix. However, when we look at the line map in Panel\,\textbf{c} of Figure\,\ref{fig:SDSSJ2359}, the \halpha\ emission from \galseven\ is very diffuse and in fact appears distinctly less bright in the central regions of the galaxy, contrary to all the other galaxies studied here. The \lya\ to \halpha\ ratio in the central 0.5 kpc radius is $12.6\pm2.9$ --- a 1.3 sigma deviation above the expected recombination ratio of $8.7$. 

We interpret this as a sign of removal of gas from the interior of the galaxy by feedback from the formation of the central star clusters. A similar situation is seen in the local starburst galaxy ESO338-IG04 where the region around one of the clusters presents a ``hole'' when viewed in both \halpha\ and [\oIII] \citep{hayes2005a, ostlin2009, bik2015}. Additionally, this evacuated region is exceptionally bright in \lya\ \citep{hayes2005a}---a situation that is very much akin to \galseven. The key difference between these cases is that for \galseven\ the vacated region is $\approx 10$ times larger than in ESO338-IG04, and while the galaxy is also quite small, this probably points to an episode of exceptionally strong feedback.

We also note that the observed E(B-V) of 0.13 (see Table\,\ref{tab: derived props}) would imply an escape fraction of $\approx30$\% for \lya, even without considering any additional effects from resonant scattering. The central escape fraction of $140\pm30$\% is clearly exceptional, at least from the central regions of a galaxy. This naturally raises the question of how  such bright \lya\ may be created without \halpha\ emission showing from the same region. There are essentially three possibilities. First, it may be the result of scattered \lya\ radiation from an outer emission region. i.e. \lya\ that is actually produced approximately cospatially with \halpha\ but that then scatters back towards the center of the galaxy. Essentially, this scenario implies that what we are seeing as a ring in \halpha\ is actually the limb brightening of a spherical shell of emitting gas. Resonant scattering of \lya\ emitted from such a shell would act as a smoothing of the structure and elevate the \lya\ to \halpha\ ratio in the center. 

Second, it may be due to collisional excitation of the \lya\ line. In principle, this mechanism can create \lya\ emission with much higher \lya\ to \halpha\ ratio than recombination. However,  the production efficiency of collisional \lya\ depends on the square of the gas density. The implication of relatively low densities in the center of the galaxy from the \halpha\ observation would then suggest that this is not a dominant mechanism. 

The third is scattering in a dusty and clumpy medium. This mechanism was initially proposed by \citet{neufeld1991} as a situation that would boost \lya\ EW. In this case, the resonant \lya\ line will scatter on the surface of the clumps and will naturally escape through low density channels while non-resonant emission will be captured and absorbed by the dense clumps. However, this scenario requires very specific gas and dust conditions to work---specifically requiring the medium to be very static relative to the \lya\ radiation, since a velocity offset between the clumps and the \lya\ photons would allow them to penetrate into the dense dusty clumps and be efficiently extincted \citep[for details see ][]{laursen2013c}.

The general behaviour of the \lya\ radial profile is very similar to \galfive, showing a inner excess rather than a classical halo. Nevertheless the emission from the extended \lya\ component is quite strong and the halo fraction of this target is $0.48\pm0.05$.

%% file: discussion.tex
\section{Discussion}\label{Discussion}

\subsection{Improving the UV modelling}\label{sec:improving the model}

The comparison with \citet{leclercq2017} used a fitting methodology that was as similar as possible to the procedure employed in that paper. However, at low-z cosmological surface brightness dimming is significantly less than at the redshifts probed by \citet[][a factor of $>200$ less when compared to redshift 4]{leclercq2017} which allows us to probe lower SB FUV emission. As can be seen from Figures~\ref{fig:SDSSJ0004}--\ref{fig:SDSSJ2359}, and Figure\,\ref{fig:model comparison} the UV light profile is not exponential beyond radii of a few kpc, but is centrally peaked within this radius and then flattens (Section~\ref{sec:improving the model}).  Similar results have already been noted by \citet{izotov2016b} and are likely attributable to a population of older stars that formed recently enough to remain bright in the UV, but are not formed in the current burst of star formation.  Even if clumpy and sparsely distributed, the light-profile of such a population may appear to decline roughly exponentially when azimuthally averaged, as done here. 

This faint, additional component would not be visible in the $z>3$ studies because cosmological surface brightness dimming would place it below the detection thresholds (see Section\ref{sec: selection methods}). These known high-z galaxies also subtend small angular sizes, and one cannot bin over very large areas to collect diffuse light.  Stacking of HST data remains a possibility to determine whether similar UV light profiles are in place at high-$z$, but is beyond the scope of this work.  
If the HST cameras cannot detect the extended continuum light but the \lya\ is captured by the much higher surface brightness sensitivity of MUSE, then halo fractions could be artificially overestimated in high-z samples that combine the two data-sets and a parametric approach.

If we want to study extended \lya\ halos and specifically the \lya\ emission that is not associated with stellar UV emission it could be important to take this faint stellar component into account. However, as we discussed in Section \ref{sec: halosizes} this additional UV emission seems to have very little effect on the \lya\ profile fits and therefore, despite the addition of a faint UV component, there is still evidence for significant extended \lya.
In two cases however, we note that the \lya\ profile is best described with a ``inner excess'' component that has an intermediate scale length, i.e. larger than the core UV component but smaller than the faint outer UV component. We interpret this as evidence that only the central bright UV component is the primary source of ionizing photons, leading to the production of extended \lya\ emission either through escape of ionizing photons---producing \lya\ in situ---or resonant scattering.

Further support for this picture comes from the age maps produced by the SED fitting routine. We measured the UV luminosity-weighted age of the region of the galaxy dominated by the inner component, and contrasted it with the age measured in an annulus around that. We find that for both targets that show an inner excess the central age is around 3 to 4 Myr, and the outer age is 600 to 700 Myr, which implies that the core is producing ionizing photons at essentially maximum efficiency, whereas the outer component is comparatively not producing any ionizing photons \citep{leitherer1999a}. The presence of an old population may bring into question the analog nature of these targets to high-z observations, but we note that the observed ages are nevertheless significantly lower than the age of the universe at the redshifts of the high-z comparison samples (z$\sim$4) and thus that such populations may exist but would be very hard to observe (see also Section\,\ref{sec:sample selection}).

\subsection{Comparison with high-z samples: selection methods and methodology}\label{sec: selection methods}
 This project was designed to image \lya\ halos in a way that allows their extended light to be quantitatively contrasted with the data from high-z studies.  The objective, however, is limited by both technology and selection effects. No FUV IFUs exist in space and we must instead use narrowband imaging.  Moreover, no current satellite has a large-volume \lya\ survey capabilities, so all targets must be pre-selected with unknown \lya\ properties, before observation with HST.

 Our galaxies were selected based on star formation and UV properties, specifically chosen to sample highly star forming galaxies even compared to the main sequence at $z\gtrsim3$, but this differs substantially from typical LBG and LAE selections. MUSE samples are a combination of UV preselection from HST imaging for sources brighter than 27th magnitude in the F775W filter and emission line selection for sources fainter than this in the HST. However, since only galaxies with significant UV continuum detections can have their sizes measured, it is likely that faint, high-EW galaxies at high-z will enter the comparison samples.  This will lessen the impact of emission line selection effects. As mentioned in Section~\ref{sec:sample selection}, our UV absolute magnitudes were selected to overlap with the fainter quartile of those in \citet{leclercq2017}, and extend to reach the magnitudes of the $L^\star$ LBGs of \citet{steidel2011}. We also show (e.g. in Figure~\ref{fig:Muse standard comparison}) that the UV scale lengths of our sample closely agree with those found for LAEs and LBGs at $z>3$, and therefore argue that differences resulting from selection should be relatively minor. We still caution, however, that these differences cannot be quantified and that there is no way to test these effects with current technology. 

We further consider the relative depths of our \lya\ observations compared to \citet{leclercq2017}, by examining the signal-to-noise ratio in the binned radial profiles as a function of surface brightness. We reach a SNR of 3 roughly at an average surface brightness of $\sim 3\times10^{-17}$ erg s$^{-1}$ cm$^{-2}$ arcsec$^{-2}$ (e.g. Panel \textbf{g} in Figures~\ref{fig:SDSSJ0004} to \ref{fig:SDSSJ2359}). In order to compare this to the results of \citet{leclercq2017} we need to take cosmological surface brightness dimming into account. For this simple comparison we assume a single redshift of 0.25 for our galaxies and 4 for the MUSE surveys, implying a difference of a factor of 256 in surface brightness.  If the faintest isophotes that we can detect at $z\approx0.25$  were instead produced by galaxies at $z=4$, the corresponding surface brightness would be $\simeq 1\times10^{-19}$ erg s$^{-1}$ cm$^{-2}$ arcsec$^{-2}$---this is indeed very similar to the detection limits given for individual galaxies in the MUSE-Deep sample \citep{leclercq2017}, and also corresponds to the faintest level shown in the full stack of LBGs by \citet{steidel2011}.  We conclude that our observations reach comparable depths in the restframe to those measured at high-redshifts. 

\subsection{Do characteristic \lya\ halo sizes evolve with redshift? }
 Despite the differences in selection, observation techniques and potential variation in observation depth our results are broadly consistent with published results from high redshifts. 
 Panel \textbf{a} of Figure\,\ref{fig:Muse standard comparison} shows that our galaxies appear to exhibit slightly smaller \lya\ to FUV scale length ratios compared to high redshift. 
 
 While the small number of our targets limits the strength of this conclusion, the effect is strong enough that warrants a discussion of what physical differences in the CGM of these galaxies could cause a drop in the observed \lya\ to FUV scale length ratio. If the neutral gas in the CGM is clumpy instead of homogeneous, \lya\ photons could escape out through the lower density channels between the clumps, requiring less spectral redistribution for an optically thin path out of the galaxy to appear. Since scattering is essentially a random walk process this would reduce the path length \lya\ photons travel before escape, lowering the scale length of the halo. Similar effects might be achieved if the velocity structure in the CGM has significantly changed between low and high redshift. Spatially resolved spectroscopic observations of the \lya\ emission from the halo may be able to provide some clues, however such observations are very challenging at low redshift. Both of these scenarios assume that the production of the \lya\ halo is dominated by scattering. 

 Another potential cause of reduction in relative \lya\ to FUV size could come from dust being present in the halos of lower redshift galaxies. It is possible that lower-z systems have had more time to produce dust and, in particular, transport this dust from the central star-forming regions to the outskirts through outflows. This could be observable with sensitive ground-based observations of \halpha\ and \hbeta, and we will examine this in a forthcoming work.
 
 However, due to the sample size the statistical effect is not very strong, and including the results of the LARS galaxies reduces the statistical significance to the point where we cannot conclude that there is a difference between the distributions.

Turning to panel \textbf{b} of Figure\,\ref{fig:Muse standard comparison} we note that the halo flux fraction distribution of our seven galaxies is statistically consistent with the high-z sample with a mean around 70 \%. In this case, adding the LARS points would indicate that halo fractions at low-z are larger than at high-z, however we note that this effect is driven mostly by low \lya-luminosity galaxies that would not be observed at high-z. These galaxies have very little central \lya-emission---leading to very high observed halo fractions. 

When taken together these observations seem to indicate a potential reduction in the relative extent of \lya\ compared to the FUV but when all low-redshift observations are taken into account this effect is not seen and the results are inconclusive. This brings us to the conclusion that despite the elapse of $\sim 10$ Gyr between the low- and high-redshift samples, there is not much change in the spatial distribution of \lya.  This conclusion holds most strongly in absolute terms (measures of the physical sizes) but also in relative terms (with respect to the ultraviolet continuum) when considering all low-z galaxies together.  That we appear to be sampling similar objects and halos to similar depths provides additional confidence in this lack of evolution.  We further consider the spectroscopic results of \citet{hayes2021a}, in which we identified no evolution in the kinematic properties of \lya\ between comparable low- and high-$z$ galaxy samples (again using HST/COS and VLT/MUSE, respectively). Taken as a whole these results are encouraging for the use of low-redshift observations in efforts to probe the processes ongoing in early universe galaxies: the ratios between bright central emission and diffuse halo emission are comparable in both redshift regimes, and observational biases such as aperture effects will be similar.  As far as \lya\ observations can say, the distributions of dust and gas that influence the \lya\ transport do not evolve strongly.

\subsection{Extended \halpha: \lya-halos produced by in situ recombinations or resonance scattering?}\label{sec: Halpha centered fits}

We now focus on the physical origin of the halos. Is the extended emission produced by scattered radiation or is it produced by recombining gas at large radii? Since \halpha\ traces the same gas that produces \lya\ we can use our \halpha\ observations to quantify this issue. The SNR of the \halpha\ data means that we cannot trace it to large radii which complicates the comparison. Nevertheless, the experiment is interesting and we therefore perform 2 component fits to the \lya\ profile entirely akin to before with the one difference that instead of constraining the central component using the FUV data we now use the \halpha\ data. We find that the difference in quality of the fits is marginal and does not warrant any strong conclusions. Within the uncertainties of the data, the core profile of \lya\ can be adequately constrained by both FUV and \halpha. However, we do note that in many of the profiles \halpha\ does seem to be slightly more extended than the FUV.

In order to definitively answer whether \lya\ is produced by in-situ recombinations or dominated by scattering we would require more sensitive or deeper data in \halpha\ that would tell us whether the \halpha\ also shows a break in the profile slope akin to the \lya\ and at which relative surface brightness. We have obtained MUSE data for this purpose and this will be the subject of a forthcoming paper.

\halpha\ emission being extended beyond the ionizing star forming knots may have several important implications for galaxy observations, most notably the escape of ionizing radiation (LyC).  LyC emission is now frequently reported in both high- and low-z galaxy samples, where it most strongly correlates with the relative strength of \lya\  \citep{steidel2018a, izotov2018c, marchi2019, flury2022}. This is intuitive because both UV radiations are absorbed by dust and atomic hydrogen, but large scatter remains on all these relations.

It is likely that a significant amount of this scatter comes from orientation effects, where LyC measurements are direct line-of-sight estimates, while others (such as \lya) are the result of reprocessed LyC radiation that has propagated in all directions.  These observations show that some \halpha\ is produced outside the stellar knots, and consequently so is a fraction of the intrinsic \lya.  It is clear, then, that even in the absence of resonance scattering, the path taken by \lya\ radiation must necessarily differ from that of the ionizing continuum.  Similar arguments can be made about nebular line diagnostics from other species such as the [O~{\sc ii}] and [O~{\sc iii}] lines, which are also used as indirect LyC diagnostics. 
Variations in the extent of these ionized regions are likely a contributing source of the scatter on the \lya--LyC relations, and a demonstration of how even in compact galaxies at $z\sim 0.25$ \citep[similar to][]{izotov2018c,flury2022} indirect diagnostic information on LyC emission is not produced cospatially with the ionizing radiation.

\subsection{Comparison with simulations}
\citet{mitchell2021} investigated a single simulated galaxy and find that emission from the galaxy ISM, i.e. resonant scattering, dominates the central region but that in-situ emission from the CGM becomes important around 7-10 kpc. This produces a halo that is qualitatively consistent with both \citet{leclercq2017} and this work.  They also note that around 50\% of the CGM emission comes from recombination which would also naturally produce extended \halpha. 

\citet{byrohl2021} examined the relative importance of different source regions for \lya\ emission in synthetic galaxies drawn from the TNG50 \citep{nelson2019a, pillepich2019} simulation of IllustrisTNG galaxies.   Their synthetic sample spanned a halo mass range of 
$10^8$~\msun\ to $10^{12}$~\msun, and could well reproduce the light profiles published in \citet{leclercq2017}.  The halo masses of our galaxies are of course unknown, but our central surface brightnesses best correspond to the more massive halos studied by \citet{byrohl2021} when corrected for cosmological surface brightness dimming. 
They distinguish between  `intrinsic' and  `processed' \lya\ emission, where the former is defined as \lya\ that has not undergone scattering outside of the emitting cell, which has a minimum size of 100~pc in the densest regions of the ISM but declines to circumgalactic regions.  They found that the intrinsic emission often dominates the centers of their galaxies but drops quickly after around 15 kpc. This is quite consistent with the picture we present here with a low surface brightness UV emission component tracing the source of \lya\ out to larger radii than has been previously observed. Our UV emission drops somewhat faster than \lya\ of the galaxies studied in \citet{byrohl2021}, however, which may be attributable to sample selection effects within the simulations. TNG50 includes galaxies that are larger than the compact galaxies  studied here and do not have to compete with observational surface brightness limits.

%% file: conclusions.tex
\section{Summary \& Conclusions}\label{Conclusions}

We have performed a new \lya\ imaging study of galaxies at redshift 0.23-0.31, with a view to accurately measuring the contributions of extended \lya\ to the overall output, and coupling this to the distribution of ultraviolet stellar light and ionized gas.  We have produced \lya\ and \halpha\ images for seven compact starbursts that are broadly comparable to those for which similar \lya\ studies have been performed at redshifts beyond 3.  We now summarize our main findings. 

\begin{itemize}
    \item We detect \lya\ emission around all galaxies.  The average escape fraction of the sample is relatively low compared to distant galaxies, with a mean $f_{\mathrm{esc}}$ of 12\%, but quite comparable to similar galaxies at low-$z$ \citep{hayes2014, henry2015a}.  
    \item The sample appears consistent with other low-z results regarding relations between \lya\ output, in terms of EWs, and escape fractions, and properties of the interstellar medium, such as dust reddening and ionization parameter. 
    \item The observed UV sizes and luminosities of our galaxies matches the properties of high-redshift galaxies well, when measurements are made in comparable ways. 
    \item The  observed \lya\ to UV size ratio of our sample ranges between 0.8 and 16.9 with a mean of $\sim5.6$, and the fraction of \lya\ that is contributed by the extended emission is 30--90\%. While the \lya\ to UV size ratio is smaller in our galaxies, the difference between low and high redshift when also considering the LARS sample is not significant. We therefore conclude that the evidence for evolution between low and high redshift is marginal. 
    \item The low-z FUV-continuum data have significantly better intrinsic surface brightness sensitivity than those for galaxies at high-z, which enables us to identify more extended stellar light. However, we find that extending  the model above to account for this additional UV component does not significantly impact the halo fraction measurements, indicating that this UV emission comes from an older, less ionizing, stellar population which does not significantly contribute to the total \lya\ emission.
    \item We are unable to firmly determine whether the \lya\ is produced in-situ or resonantly scattered into the halo. However, we show intriguing hints that \halpha\ is more extended than the central ionizing FUV profile. This could indicate that ionizing radiation is travelling significant distances from the source before being absorbed,  implying that ionizing radiation is not entirely confined by the ISM in these galaxies. Further investigation of this with deeper \halpha\ data is underway using VLT/MUSE observations.

\end{itemize}

